\documentclass[iop]{emulateapj}
\usepackage{latexsym}
\usepackage{dcolumn}
\usepackage{bm}
\usepackage{amsmath}
\usepackage{natbib}
\input{epsf}

\newcommand{\beq}{\begin{equation}}
\newcommand{\eeq}{\end{equation}}
\newcommand{\bea}{\begin{eqnarray}}
\newcommand{\eea}{\end{eqnarray}}
\newcommand{\bi}{\begin{itemize}}
\newcommand{\ei}{\end{itemize}}
\newcommand{\bfi}{\begin{figure}[!t]
\epsfxsize=9cm
\epsffile}
\newcommand{\bfig}{\begin{figure*}[!t]
\center{}
\epsfxsize=15cm
\epsffile}
\newcommand{\efi}{\end{figure}}
\newcommand{\efig}{\end{figure*}}
\newcommand{\no}{\nonumber}
\newcommand{\mpch}{{\rm Mpc}/h}
\newcommand{\hmpc}{h/{\rm Mpc}}
\newcommand{\bfk}{{\bf k}}
\newcommand{\bfv}{{\bf v}}
\newcommand{\tauT}{\tau_{\rm T}}
\newcommand{\bfs}{\bf s}

\begin{document}
\title{Robustness of the pairwise kinematic Sunyaev-Zeldovich power spectrum shape as a cosmological gravity probe}
\author{Yi Zheng}
\email[Email me at: ]{zhengyi27@mail.sysu.edu.cn}
\affiliation{School of Physics and Astronomy, Sun Yat-sen University, 2 Daxue Road, Tangjia, Zhuhai, 519082, China}

\begin{abstract}
We prove from the modified gravity (MG) galaxy/halo mock catalogs that the shape of the pairwise kinematic Sunyaev-Zeldovich (kSZ) power spectrum $P_{\rm kSZ}$ has constraining power on discriminating different gravity theories on cosmological scales. By varying the effective optical depth $\tau_{\rm T}$ as a free parameter, we verify that the $\tau_{\rm T}$-$f$ (the linear growth rate) degeneracy in the linear theory of $P_{\rm kSZ}$ is broken down by the non-linear structure growth and the scale-dependence of $f$ in some MG theories. Equivalently speaking, the shape of $P_{\rm kSZ}$ alone could be used to constrain the MG theories on cosmological scales. With a good knowledge of galaxy density biases, we verify that, a combination of the next generation of CMB and galaxy spectroscopic redshift surveys, e.g. BOSS+CMB-S4 or DESI+CMB-S4, could potentially discriminate $f(R)$ models from the general relativity at $\sim5\sigma$ level using the shape of the galaxy pairwise kSZ dipole $P_{{\rm kSZ},\ell=1}$ alone, when $f_{R0}=10^{-4}$.
\end{abstract}
\pacs{98.80.-k; 98.80.Es; 98.80.Bp; 95.36.+x}
\maketitle


\section{Introduction}
\label{sec:intro}

The cosmological test of gravity theories has gain much more attention since the discovery of the cosmic acceleration. Profound progresses  have been made, though there is still room to explore from the next generation of cosmological surveys, which will collect at least one order of magnitude more data than all the existing ones in the current literature. Therefore many next-to-leading-order cosmological/astrophysical effects will be detected at a high confidential level and will significantly benefit the cosmological gravity test. The kinematic Sunyaev-Zeldovich (kSZ) effect \citep{kSZ1970,kSZ1972,kSZ1980,kSZ1986} is one of these.

The kSZ effect describes one of the secondary CMB temperature anisotropies, induced by the CMB photons Thomson scattering off a bunch of free electrons with bulk motion. It is a measure of both the cosmological velocity field and baryon distribution in the universe, and the induced CMB temperature change
\bea
\delta T_{\rm kSZ}(\hat{n}) &=& -T_0\int dl\sigma_{\rm T} n_e \left(\frac{{\bfv}_e\cdot \hat{n}}{c}\right)  \, ,\\
\delta T_{\rm kSZ}(\hat{n}_i) &=& - \frac{T_0\tau_{{\rm T},i}}{c}\bfv_i\cdot\hat{n}_i \,.
\label{eq:T_i}
\eea
Here $T_0\simeq 2.73\rm K$ is the averaged CMB temperature, $\sigma_{\rm T}$ is the Thomson scattering cross-section, $c$ is the speed of light, $\hat{n}$ is the unit vector along the line-of-sight (LOS), $n_e$ is the physical free electron number density, $\bfv_e$ is the peculiar velocity of free electrons, and the integration $\int dl$ is along the LOS given by $\hat{n}$. Eq. (\ref{eq:T_i}) assumes that the CMB photons scatter off only one cloud of moving free electrons surrounding the $i$th halo until they reach the observer, and $\tau_{{\rm T},i}=\int dl\sigma_{\rm T} n_{e,i}$ is the optical depth of the $i$th halo. We adopt this assumption through out the paper. Furthermore, we assume a constant $\tauT$ for all halos for simplicity, thus $\tauT$ could be singled out in all calculations.

The velocity field of free electrons is believed to be a good tracer of dark matter velocity field and hence a promising test bed of cosmological models. Several works have detected the kSZ effect in the literature, although the signal-to-noise (S/N) has not reached $5\sigma$ \citep{Hand12,Soergel16,Schaan16,PlanckkSZ16,Hill16,DeBernardis17,Sugiyama18,PlanckkSZ18}. By combining the next generation of galaxy redshift and CMB data, the significance of kSZ detection could potentially reach $20-100\sigma$, by varying scenarios \citep{Sugiyama18,Flender16}. This high detection significance will heavily benefit the constraint on the DE properties and the MG theories \citep{Sugiyama2017}.

Though promising, this field always suffers from our poor understanding of $\tauT$. The calculation of $\tauT$ is based on various complicated astrophysical processes, and could be only possible in hydro-dynamical simulations, which still suffers from numerous uncertainties nowadays. As one could tell from Eq. (\ref{eq:T_i}), $\tauT$ is degenerated with the amplitude of velocity field. Therefore the uncertainty of $\tauT$ will heavily degrade the constraining power of kSZ effect on cosmological models. Additionally, several systematic errors in the  kSZ detection, such as the miscentering bias and scatter-in-mass bias, will systematically decrease the overall amplitude of kSZ signal \citep{Flender16} and induce systematic biases if we rely on the amplitude of kSZ signal to constrain cosmology. Consequently, the conservative choice is to merely use the shape of the kSZ signal for cosmological analysis. This work will prove the robustness of this idea using high resolution MG N-body simulations. 

\begin{table*}
\center{
\scalebox{1}{
\begin{tabular}{cccccccc}
\hline\hline
FWHM & noise & galaxy & redshift & $V$ & $\bar{n}_{\rm g}$  & $M_{\rm avg}$ \\
$[{\rm arcmin}]$ & $[\mu K\mathchar`-{\rm arcmin}]$ &  &  & $[(h^{-1}\,{\rm Gpc})^3]$ & $[(h^{-1}\,{\rm Mpc})^{-3}]$ &  $[h^{-1}\, M_{\sun}]$ \\

\hline
\hline
CMB-S4 + BOSS\\
\hline
$1$ & $2$ & LRG & 0.5 $(0.43<z<0.70)$ & 4.0 & $3.2\times10^{-4}$ & $2.6\times10^{13}$ \\
\hline
CMB-S4 + DESI\\
\hline
$1$ & $2$ & LRG & 0.8 $(0.65<z<0.95)$ & 13.5 & $2\times10^{-4}$ & $2.6\times10^{13}$ \\
\hline
\end{tabular}
}}
\caption{The assumed survey specifications for forecasts. The first two columns show beam size and detector noise in a CMB-S4-like CMB experiment. We assume an ideal beam size $1\arcmin$. From the third to seventh columns show the type of galaxies, redshift $z$, comoving survey volume $V$, mean comoving number density $\bar{n}_{\rm g}$, and the averaged halo mass $M_{\rm avg}$.}
\label{table:survey}
\end{table*}

\section{The pairwise kSZ power spectrum}
\label{sec:P_kSZ}

Halos containing free electrons tend to move toward each other, due to the mutual gravitational attractions. This peculiar kinematic pattern leaves distinct feature in the CMB map, which could be captured by the pairwise kSZ estimator in the Fourier space \citep{Ferreira1999,Sugiyama18},
\bea
\label{eq:P_kSZ}
P_{\rm kSZ}(\bfk) &=& \left\langle -\frac{V}{N^2} \sum_{i,j}\left[\delta T_{\rm kSZ}(\hat{n}_i)-\delta T_{\rm kSZ}(\hat{n}_j)\right]e^{-i \bfk\cdot {\bfs}_{ij}} \right\rangle \, \no \\ 
& \simeq & \left(\frac{T_0\tauT}{c}\right) P_{\rm pv}(\bfk) \,, \\
\label{eq:P_pv}
P_{\rm pv}(\bfk) &=& \left\langle \frac{V}{N^2} \sum_{i,j}\left[{\bfv}_i \cdot \hat{n}_i-{\bfv}_j \cdot \hat{n}_j\right]e^{-i \bfk\cdot {\bfs}_{ij}} \right\rangle \,.
\eea
$P_{\rm kSZ}$ is the pairwise kSZ power spectrum in redshift space, $P_{\rm pv}$ is the galaxy LOS pairwise velocity power spectrum in redshift space, $V$ is the survey volume, $N$ is the number of galaxies, and ${\bfs}_{ij}={\bfs}_i-{\bfs}_j$ is the galaxy separation vector in redshift space.

In a series of papers \citep{Sugiyama2016,Sugiyama18,Sugiyama2017}, Sugiyama et al. proved that, assuming the global plane-parallel approximation $\hat{n}_i \sim \hat{n}_j \sim \hat{n}$ and the amplitude of velocity field is proportional to $f$, we could drive $P_{\rm pv}(\bfk)$ from the redshift space galaxy density power spectrum $P_{s}(\bfk)$ in the following way,
\beq
P_{\rm pv}(\bfk) = \left(i\frac{aHf}{\bfk \cdot \hat{n}}\right) \frac{\partial}{\partial f} P_{\rm s}(\bfk) \,,
\label{eq:P_kSZ_der}
\eeq
where $a$ is the scale factor, $H$ is the Hubble parameter at redshift $z$. This relation holds for any object such as dark matter particles, halos, galaxies, and galaxy clusters with anisotropic clustering property (in redshfit space).

Instead of detailed modelling of $P_{\rm kSZ}$ \citep{Sugiyama2016,Okumura2014}, we provide here a toy model to qualitatively understand the numerical results. We introduce a simple RSD model, with a linear Kaiser term and a Gaussian Finger-of-God (FoG) term,
\beq
P_{\rm s}(k,\mu) = \left(b+f\mu^2\right)^2P_{\rm lin}(k)\exp(-k^2\mu^2\sigma_v^2/H^2)\,.
\label{eq:simple_rsd}
\eeq
Here $b$ is the linear galaxy bias, $\mu$ denotes the cosine of the angle between $\bfk$ and the LOS, and $P_{\rm lin}(k)$ is the linear dark matter power spectrum at redshift $z$. In linear theory, the LOS velocity dispersion $\sigma_v^2=\int f^2H^2P_{\rm lin}(k)dk/6\pi^2$. By substituting Eq. (\ref{eq:simple_rsd}) into Eq. (\ref{eq:P_kSZ_der}), we derive the corresponding formula for $\Delta_{\rm kSZ}(k,\mu)\equiv k^3P_{\rm kSZ}(k,\mu)/2\pi^2$,
\bea
\label{eq:P_kSZ_toy}
\Delta_{\rm kSZ}(k,\mu)&=&\frac{k^3}{2\pi^2}\left(\frac{T_0\tau_{\rm T}}{c}\right)2iaHf\mu(b+f\mu^2)\frac{P_{\rm lin}(k)}{k} \no \\
&&\times  S(k,\mu) \exp(-k^2\mu^2\sigma_v^2/H^2)\,, \\
S(k,\mu)&=&1-(b/f+\mu^2)k^2\sigma_v^2/H^2\,.
\eea

Eq. (\ref{eq:P_kSZ_toy}) characterizes two circumstances at which the $\tau_{\rm T}-f$ degeneracy could be broken. One is when the MG theory predicts a scale-dependent $f$, which will change the shape of $\Delta_{\rm kSZ}$ in a characteristic way and break the $\tau_{\rm T}-f$ degeneracy. The other is the non-linear structure growth, which generates higher order terms such as the shape kernel $S(k,\mu)$ and the FoG term. The combination of the term $\tauT fb$ and these higher order terms will also break the $\tauT-f$ degeneracy.

In order to illustrate the robustness of above two mechanisms in breaking the $\tauT-f$ degeneracy, and also to quantify the robustness of the $\Delta_{\rm kSZ}$ shape as a cosmological gravity probe, we will measure and compare the $\Delta_{\rm kSZ}$ dipoles from high-resolution general relativity (GR) and MG simulations in the following sections.

\section{The MG simulations and mock catalogs}
\label{sec:simu}

 We study two representative MG theories in this work, the $f(R)$ gravity \citep{fRreview} and the normal branch of DGP (nDGP) gravity \citep{DGP}. The $f(R)$ gravity acquires a scale-dependent $f$ and we adopt the Hu $\&$ Sawicki (HS) functional form of $f(R)$, where the deviation from GR is characterized by the free parameter $|f_{R0}|=|\partial f/\partial R|_{z=0}$. The nDGP model predicts a scale-independent $f$, and this model has one parameter $r_c$ of length dimension, below which gravity becomes four dimensional.

$f(R)$ simulations are run by the ECOSMOG code \citep{ECOSMOG,Bose2017} and nDGP simulations are run by the ECOSMOG-V code \citep{Li2013,Barreira2015}. The simulation box size is $1024\mpch$ and the particle number is $1024^3$. Three $|f_{R0}| = 10^{-4}, 10^{-5}, 10^{-6}$ values are chosen for f(R) gravity simulations, denoted as F4, F5, and F6. Two $H_0r_c = 1.0, 5.0$ values are adopted for nDGP gravity simulations, named as N1 and N5. The levels of deviation from GR are in the sequence of F4$>$F5$>$F6, and N1$>$ N5. All simulations have the same background expansion quantified by the WMAP9 cosmology \citep{WMAP9},
\beq
\{\Omega_{\rm b}, \Omega_{\rm CDM}, h, n_s, \sigma_8 \} = \{0.046, 0.235, 0.697, 0.971, 0.82\} \,, \no
\eeq
and we run 5 realizations for each gravity model.

The $z=0.5$ and $z=0.8$ snapshots are analyzed in this work, mimicking two galaxy catalogs, being respectively the CMASS sample of the BOSS survey \citep{Manera2013} and the LRG galaxy sample of the DESI survey from $z= 0.65$ to $z=0.95$ \citep{DESI16I, Sugiyama18}. The dark matter halo catalogs are generated using ROCKSTAR \citep{Rockstar}. At $z=0.5$, by tuning the HOD parameters suggested in \citep{ZhengZheng2007}, the mock galaxy catalogs are generated. Different HOD parameters are applied to different MG simulations, and all ``constrained'' galaxy catalogs are guaranteed to have the identical galaxy number density $n_g$ and the projected correlation functions $w_p(r_p)$. This effectively fixes the uncertainties from the complicated galaxy density biases and we could therefore focus solely on the physical deviations induced by the MG theories at cosmological scales. The HOD parameters of the GR simulations are the best-fit HOD parameters from the CMASS data \citep{Manera2013}. At $z=0.8$, we simply select all dark matter halos with $M>10^{13}M_\cdot/h$ to represent the LRG galaxies of the future DESI survey from $z=0.65$ to $z=0.95$, for the purpose of a rough $S/N$ estimation. The detailed description of the simulations and catalogs could be found in \citep{Cesar2019}.

For real to redshift space, we use the following formula to move the positions of the mock galaxies/halos:
\beq
\label{eq:mapping}
{\bf s}={\bf r}+\frac{{\bf v} \cdot \hat{n}}{a(z)H(z)}\hat{n}\,,
\eeq
where {\bf r} is the real space position and {\bf s} is the redshift space position.

Besides the mock catalogs, the signal-to-noise ratio prediction also depends on the specific survey parameters. We choose a CMB-S4 like survey as our CMB survey baseline \citep{CMBS42019}, and choose two galaxy survey setups, the BOSS-like and DESI-like surveys as redshift survery baselines. The detailed survey specifications are shown in Table \ref{table:survey}.

\bfi{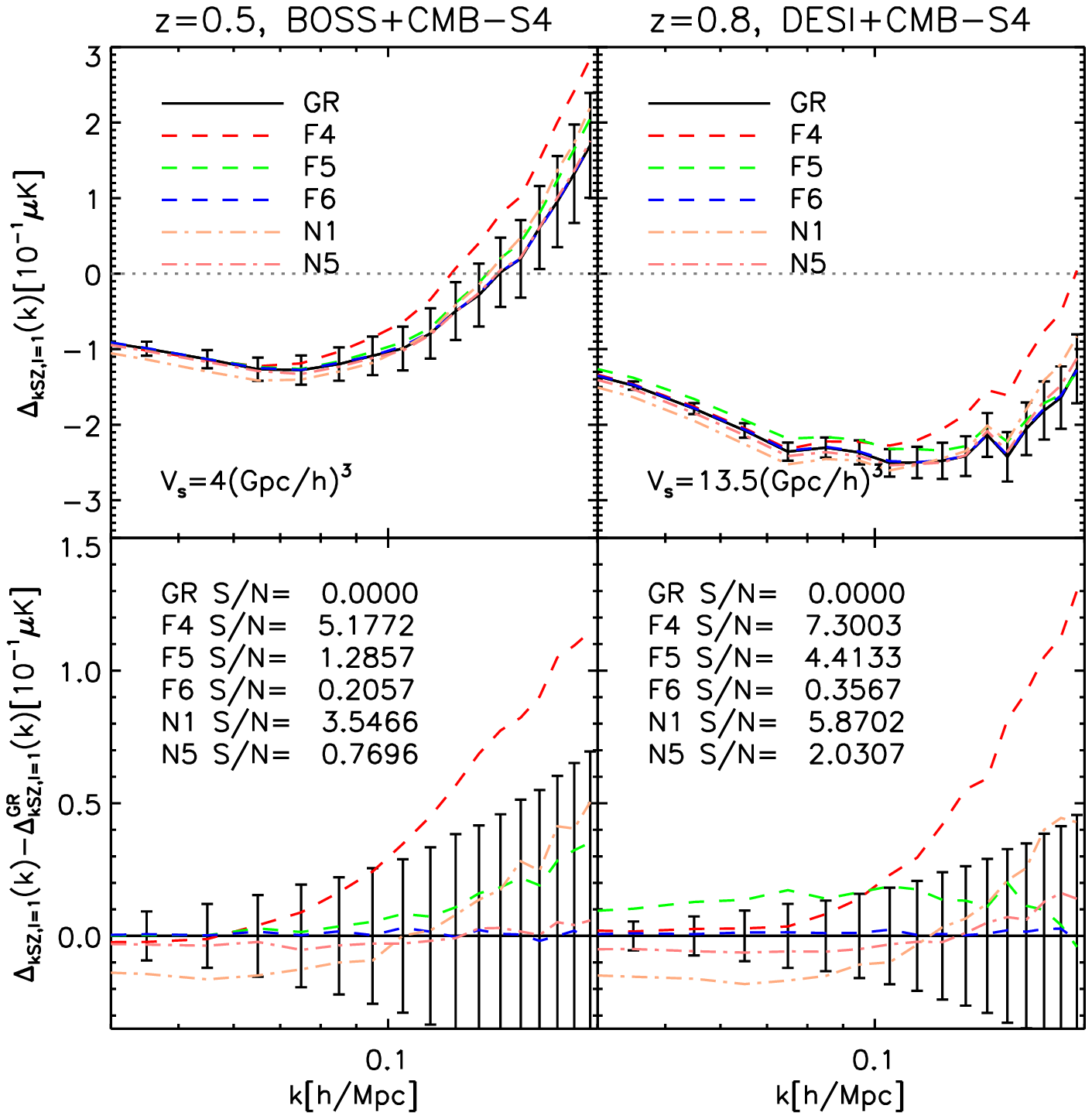}
\caption{ Top panels: The measured dimensionless dipoles $\Delta_{{\rm kSZ},\ell=1}= i^{\ell=1}k^3P_{{\rm kSZ}, \ell}(k)/2\pi^2$ of CMB-S4+BOSS-like and +DESI-like mock galaxies. Bottom panels: the dipole differences between the MG and GR simulations $\Delta_{{\rm kSZ}, \ell=1}-\Delta_{{\rm kSZ}, \ell=1}^{\rm GR}$. For conciseness, we only plot the error bars of GR simulation for illustration.}
\label{fig:ksz_mg_fix_tau}
\efi
\bfig{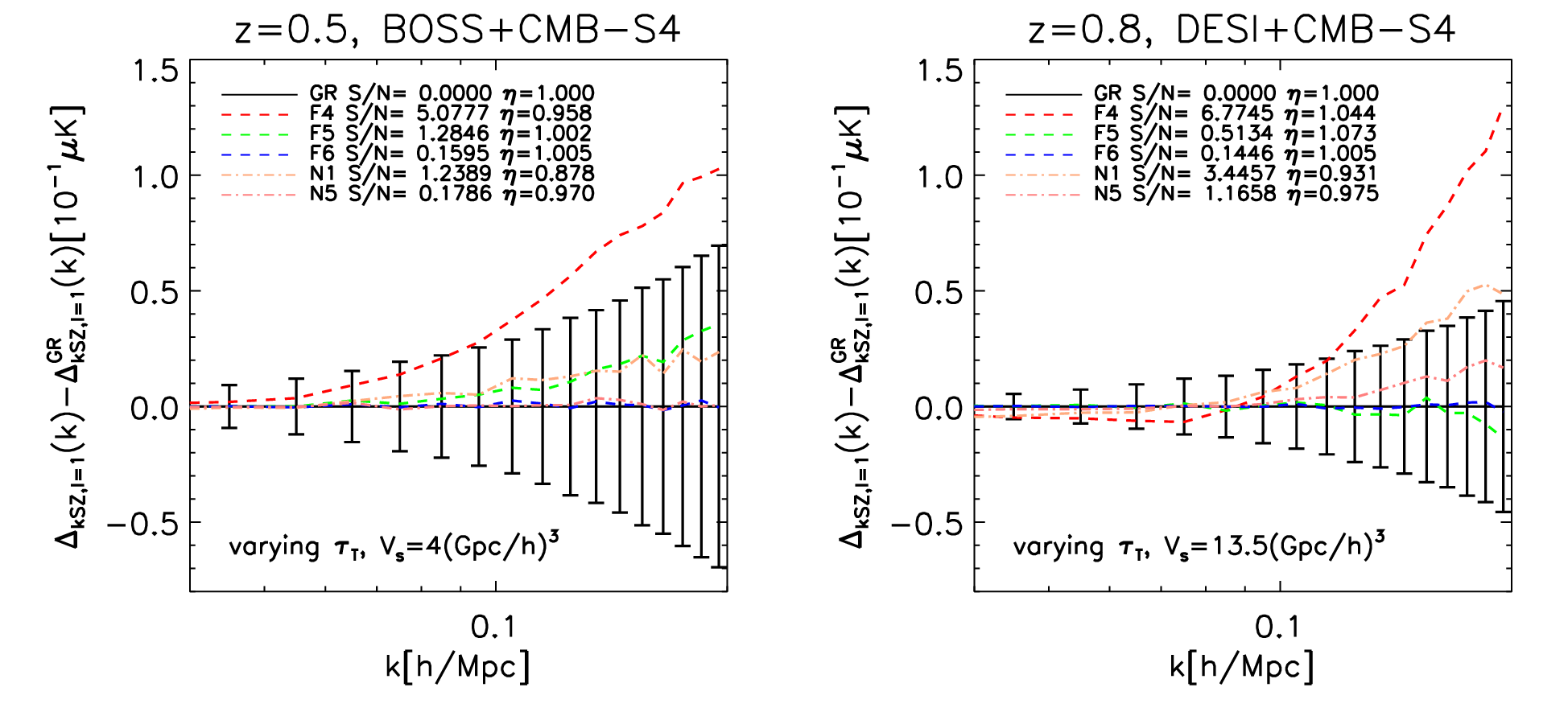}
\caption{Similar to the bottom panels of Fig. \ref{fig:ksz_mg_fix_tau}, but with a varying $\tauT$.}
\label{fig:ksz_mg_varying_tau}
\efig
\section{The pairwise kSZ dipole}
\label{sec:dipole}

We study the dipole of $\Delta_{\rm kSZ}(k,\mu)$ in this section. The first step is to calculate $P_{\rm pv}(k,\mu)$. For the computational convenience, we adopt an estimator equivalent to Eq. (\ref{eq:P_pv}), namely \citep{Sugiyama2016}
\beq
(2\pi)^3\delta_{\rm D}(\bfk+\bfk')P_{\rm pv}(\bfk)=\left<p_{\rm s}(\bfk)\delta_{\rm s}(\bfk')-\delta_{\rm s}(\bfk)p_{\rm s}(\bfk')\right> \,,
\eeq
where $p_{\rm s}({\bfs})=[1+\delta_{\rm s}({\bfs})][\bfv({\bfs})\cdot \hat{n}]$ and $\delta_{\rm s}({\bfs})$ are respectively the momentum and density fluctuation fields in redshift space. $p_{\rm s}(\bfk)$ and $\delta_{\rm s}(\bfk)$ are their corresponding Fourier counterparts. We sample the $p_{\rm s}(\bfs)$ and $\delta_{\rm s}(\bfs)$ fields on $1024^3$ regular grids using the nearest-grid-point (NGP) method and calculate $p_{\rm s}(\bfk)$ and $\delta_{\rm s}(\bfk)$ fields by the Fast Fourier transform (FFT) method.

By choosing the AP filter radius $\theta_{\rm c}$ maximizing the signal-to-noise\footnote{The methodology is detailed in the appendix \ref{app-sec:cov}}., we derive the average $\tau_T$'s of two target galaxy catalogs. At $z=0.5$, $\tau_T = 4.5\times10^{-5}$, $\theta_{\rm c} = 1.37'$, and at $z=0.8$, $\tau_T = 6.0\times10^{-5}$, $\theta_{\rm c} = 1.12'$.

Therefore, Eq. (\ref{eq:P_kSZ}) gives $P_{\rm kSZ}(k,\mu)$. Using the Legendre polynomials $\mathcal{P}_\ell(\mu)$, the multipole of $P_{\rm kSZ}(k,\mu)$ is defined as
\beq
P_{{\rm kSZ}, \ell}(k) = \frac{2\ell+1}{2}\int_{-1}^1 d\mu P_{\rm kSZ}(k,\mu)\mathcal{P}_\ell(\mu) \,.
\label{eq:P_kSZ_multipole}
\eeq
The measured dimensionless dipoles $\Delta_{{\rm kSZ},\ell=1}= i^{\ell=1}k^3P_{{\rm kSZ}, \ell}(k)/2\pi^2$ of CMB-S4+BOSS-like and +DESI-like mock galaxies are shown in the top panels of Fig. \ref{fig:ksz_mg_fix_tau} \footnote{We also calculate and compare the octopoles of $P_{\rm kSZ}$, but find negligible differences ($<1\sigma$) of $P_{{\rm kSZ},\ell=3}$ between different gravity simulations.}.

We see that all simulations give the $\Delta_{{\rm kSZ},\ell=1}$ of similar trend, with a turnover at around $k\sim0.08\hmpc$. However different gravity theories predict $\Delta_{{\rm kSZ},\ell=1}$ of different amplitudes and shapes, and the MG dipoles are more distorted than that of GR. The deviations of F4 and N1 simulations from GR case are larger than that of F5/F6 and N5 simulations, as expected. 

Instead of making careful comparisons between Eq. (\ref{eq:P_kSZ_toy}) and Fig. \ref{fig:ksz_mg_fix_tau}, we generally discuss the impact of the shape kernel $S(k,\mu)$ on the dipole shape here. $S(k,\mu)$ is physically a consequence of the competition between Kaiser and FoG effects, which makes it a decreasing function ranging from unity to $-\infty$ as $k$ increases. Together with the FoG term, $S(k,\mu)$ induces the turnover of $\Delta_{{\rm kSZ},\ell=1}$ at around $k\sim0.08\hmpc$. Furthermore, as $k$ crosses a typical $k_{S=0}(\mu)=\sqrt{3H^2/[(b/f+\mu^2)\sigma_v^2]}\propto(f/b+f^2\mu^2)^{-1/2}$ for a given $\mu$, $\Delta_{{\rm kSZ}}(k,\mu)$ value will transfer from negative to positive, and continues to grow as Fig. \ref{fig:ksz_mg_fix_tau} illustrates \footnote{$k_{S=0}\propto(f/b+f^2\mu^2)^{-1/2}$, thus gravity generating larger $f$ will predict smaller $k_{S=0}$. Galaxies have $b>1$, thus their $k_{S=0}$ will be smaller than that of dark matter. Both predictions are consistent with Fig. \ref{fig:ksz_mg_fix_tau}. Moreover, it might be possible to use $k_{S=0}$ to constrain $f$, which is beyond the scope of this paper.}. Finally, different gravity theories predict different $f$, $P_{\rm lin}(k)$, $\sigma_v^2$ and $b$. These physical quantities interact together in Eq. (\ref{eq:P_kSZ_toy}) and generate diverse kSZ dipoles shown in Fig. \ref{fig:ksz_mg_fix_tau}.

In the bottom panels of Fig. \ref{fig:ksz_mg_fix_tau}, we show the dipole differences between the MG and GR simulations $\Delta_{{\rm kSZ}, \ell=1}-\Delta_{{\rm kSZ}, \ell=1}^{\rm GR}$. The nDGP model predicts a constant $f$ deviating from GR case at all scales, while the $f$ of $f(R)$ model coincides with that of GR at large scales and then increases towards smaller scales \citep{Cesar2019}. Therefore, at large scales (e.g., $k \lesssim 0.08\hmpc$), the deviations of nDGP dipoles are in general larger than those of $f(R)$ models and in the other way around at small scales (e.g., $k \gtrsim 0.1\hmpc$). 

We calculate the diagonal elements of the covariance matrix of the dipole measurements by Eq. (\ref{eq:cov}). Other than the shot-noise of  galaxy distribution, the primary CMB anisotropies and the detector noise of the CMB experiment are considered as well. Then we evaluate the S/N of the differences between MG and GR dipoles, where 
\beq
{\rm \frac{S}{N}}=\sqrt{\chi^2}=\sqrt{\frac{\sum_i\left[\Delta_{{\rm kSZ},\ell=1}(k_i)-\Delta_{{\rm kSZ},\ell=1}^{\rm GR}(k_i)\right]^2} {\sigma^2_{\Delta}(k_i)}}\,.
 \label{eq:s_to_n}
\eeq
$k_{\rm min}=0.035\hmpc$, $k_{\rm max}=0.195\hmpc$ and $\Delta k=0.01\hmpc$.

The estimated S/N are listed in the bottom panels of Fig. \ref{fig:ksz_mg_fix_tau}. We find that, combined with the CMB-S4 like CMB survey, both BOSS- and DESI-like could discriminate F4 model from GR at a $\sim 5\sigma$ level using $\Delta_{{\rm kSZ},\ell=1}$ alone. DESI-like survey, due to its large survey volume, in general has smaller error bars and higher S/N. Therefore, the pairwise kSZ power spectrum could help improving the capacity of gravity test of the next generation CMB and galaxy surveys.

\section{$\tau_{\rm T}-f$ degeneracy breaking}
\label{sec:degeneracy}

In the realistic data analysis, it is however difficult to have accurate knowledge of $\tauT$ in advance and thus the amplitude of kSZ power spectrum could not be used to constrain cosmology. The key question to ask is that, ``without knowing $\tauT$, can we still discriminate kSZ signals predicted by different gravity theories?'' The answer is yes, that we could discriminate them using the shape of the galaxy pairwise kSZ dipole.

In order to illustrate this point, we implement the following test. We fix the GR $\Delta_{{\rm kSZ},\ell=1}$ as it is in the previous section. Then in calculating the MG dipoles, we replace $\tauT$ by $\eta\tauT$, and vary $\eta$ between $[0.5,1.5]$ to fit the GR dipole. Consequently we obtain the best fitted $\eta$, and the corresponding S/N of the GR-MG dipole differences are calculated by Eq. (\ref{eq:s_to_n}). We consider this S/N as the constraining power on the gravity models from the shape of the galaxy pairwise kSZ dipole, and equivalently, it illustrates how much the $\tauT-f$ degeneracy is broken by the $\Delta_{{\rm kSZ},\ell=1}$ shape. The results are plotted in Fig. \ref{fig:ksz_mg_varying_tau}.

It is shown that, (1) in nDGP cases, the amplitudes of $\Delta_{{\rm kSZ},\ell=1}$ change by up to$\sim10\%$, in $f(R)$ cases, the amplitudes change by up to $\sim4\%$, (2) the S/N decreases compared with the fixed $\tauT$ case, but with a minor level. This small discrepancy illustrates the facts that (1) non-linear structure growth already breaks the $\tauT-f$ degeneracy in the kSZ signal, and (2) the scale-dependence of $f$ in $f(R)$ models further breaks the $\tau-f$ degeneracy as discussed previously. Therefore, we prove that using the $\Delta_{{\rm kSZ},\ell=1}$ shape alone, we could discriminate F4 gravity from GR at $\sim5\sigma$ level, by a BOSS+ or DESI+CMB-S4 like combination. Moreover it could contribute to the constraint of F5, N1, and N5 like gravity theories in the future. This conclusion overcomes the obstacle in the kSZ cosmology originating from the poor understanding of $\tauT$ and will have a potential implication prospect in the next generation of galaxy and CMB surveys, complementing other cosmic probes in constraining dark energy and gravity theories.

In the analysis, we choose a moderate $k_{\rm max}=0.195\hmpc$ in the analysis, which corresponds to a comoving scale around $5.1\hmpc\sim32.2\hmpc$. Therefore our conclusions are immune from the systematics affecting the small scale kSZ signal, e.g.,  the complicated small scale astrophysical processes.

\section{Discussions}
\label{sec:discussion}

In conclusion, we verify that the shape of the galaxy pairwise kSZ dipole has potential constraining power on gravity models at cosmological scales. This constraining ability does not depend on the small scale kSZ signal, thus immune from the complicated small scale astrophysical processes. This probe is in particular useful for a self-consistent test of GR from cosmological data, where our main target is to falsify GR rather than to determine the ``true'' gravity model. We find that, with a good knowledge of galaxy density biases, a BOSS+ or DESI+CMB-S4 like survey combination could discriminate F4 gravity models from GR at $\sim5\sigma$ level, and could also contribute to the constraint of F5, N1, and N5 like gravity theories in the future, illustrating the promising implication potential of $\Delta_{\rm kSZ}$ shape in the future.

As a proof-of-concept paper, this work is simplified and could be improved in several aspects. 

(1) It would be beneficial to develop more accurate analytical models or numerical emulators for $\Delta_{\rm kSZ}(k,\mu)$. 

(2) We assume that we have good knowledge of galaxy density bias in this work, while in reality the biases or HOD parameters fitted from $w_p$ will have uncertainties. 

(3) Current forecasts assume a Gaussian gas profile and a top-hat aperture photometry filter. The real signal could be larger than current prediction if the real gas profile is more compact than Gaussian distribution and if we adopt a matched filtering technique \citep{Alonso2016} . 

(4) To obtain better forecasts of the next generation galaxy and CMB surveys, the realistic kSZ catalogs matching their survey designs would be necessary. 

(5) In a given galaxy sample, we assume that all halos have the same optical depth, and that there is no redshift dependence. This is a rough assumption. As long as the optical depths of halos do not correlate with each other on the scales that we are interested in, this assumption is robust at least on the leading order and we could consider the `effective’ optical depth as the average of the optical depths in a halo mass bin.

(6) We work at a single snapshot with a fixed redshift in this work. In the real data analysis, It requires a robust estimation of an ‘effective’ redshift of the galaxy sample if we assume a redshift independent growth function in our data analysis. Given that reasonably large volume surveys are considered and  the growth function will vary along each line of sight, this is an important topic in both galaxy RSD analysis and kSZ analysis. Without this, we will introduce systematic error to our final growth function estimation.

For comparison, The SZ tomography technique\citep{Pan2019,Smith2018} well incorporates the halo mass dependence of optical depth and redshift dependent growth function by dividing the given galaxy sample into fine enough redshift bins and halo mass bins. The same spirit has been adopted in the galaxy RSD analysis \citep{ZhengJ2018} and could be applied in our future kSZ analysis. 

Finally, we would like to use this work to motivate more theoretical and observational kSZ studies on the cosmological model constraints in the future.

\section{Acknowledgements}
We appreciate Baojiu Li and Wojciech Hellwing for providing the simulation and catalog data. We thank the anonymous referee for valuable comments that improve this paper. We thank Pengjie Zhang, Baojiu Li, Naonori S. Sugiyama, Weiguang Cui and Ziyang Chen for useful discussions. This project has also benefited from numerical computations performed at the Interdisciplinary Centre for Mathematical and Computational Modelling (ICM) University of Warsaw  under grants no GA67-17 and GA65-30.

\appendix
\section{$\tau_T$ and covariance matrix modelling}
\label{app-sec:cov}
We adopt the following formula to calculate $\tau_{\rm T}$ \citep{Sugiyama18},
\beq
\tau_T = \frac{\sigma_T f_{\rm gas}M_{\rm avg}}{\mu_{\rm e}m_{\rm p}D^2_{\rm A}(z_{\rm eff})}\int \frac{{\rm d}^2\ell}{(2\pi)^2}U(\ell \theta_{\rm c})N({\bf \ell})B({\bf \ell})\,.
\label{eq:tau_T}
\eeq
Here $f_{\rm gas}$ is the gas-mass fraction, $f_{\rm gas} = f_{\rm b}$ and $f_{\rm b}=\Omega_b/\Omega_m=0.155$ is the universal baryon fraction \citep{PLANK2015}. $M_{\rm avg}$ is the averaged halo mass in the targeted halo mass bin, calculated from simulations. $\mu_e= 1.17$ is the mean particle weight per electron, and $m_p$ is the proton mass. $D_A(z_{\rm eff})$ is the angular diameter distance at the effective redshift. $U(\ell \theta_{\rm c})$ is the Fourier transform of the AP filter \citep{Sugiyama18,Alonso2016} and $\theta_{\rm c}$ is the AP filter radius. $N(\ell)$ is the Fourier transform of the projected gas profile which we assume Gaussian. $B(\ell)$ is the Fourier transform of the Gaussian Planck beam function. 

Considering only the diagonal Gaussian terms, the covariance matrix of $P_{{\rm kSZ},\ell}$ is \citep{Sugiyama2017}
\bea
\label{eq:cov}
{\rm Cov}(P_{{\rm kSZ},\ell_1}(k), P_{{\rm kSZ},\ell_2}(k)) &&=\frac{1}{N_{\rm mode}(k)}\frac{2(2l_1+1)(2l_2+1)}{4\pi} \int d\varphi \int d\mu {\cal L}_{\ell_1}(\mu){\cal L}_{\ell_2}(\mu) \no\\
&&\times \left(\frac{T_0\tau_T}{c}\right)^2\left[2\left(P_{\rm s}^{(1)(0)}(\bfk)\right)^2-2\left(P_{\rm s}^{(1)(1)}(\bfk)+(1+R_{\rm N}^2)\frac{\sigma^{(2)}_{\rm v}}{\bar{n}}\right)\left(P_{\rm s}(\bfk)+\frac{1}{\bar{n}}\right)\right]\,.
\eea
Here $N_{\rm mode}(k)$ is the number of $k$ modes in a given $k$ bin.${\cal L}_{\ell}(\mu)$ is the  Legendre polynomials. $\bar{n}$ is the number density of mock galaxies or halos. $\sigma_{\rm v}^{(2)}$ is the dispersion of density weighted velocity of mock galaxies or halos, measured from simulations. In redshift space, $P_{\rm s}^{(1)(0)}$ is the momentum-density cross power spectrum, $P_{\rm s}^{(1)(1)}$ is the momentum-momentum auto power spectrum, and $P_{\rm s}$ is the density-density auto powers spectrum. All three power spectra are calculated from simulations.

In particular, we define the inverse signal-to-noise ratio $R_{\rm N}$
\beq
R^2_{\rm N}(\theta_{\rm c}) = \frac{\sigma^2_{\rm N}(\theta_{\rm c})}{\sigma_{\rm kSZ}^2(\theta_{\rm c})}\,,
\eeq
with the signal
\beq
\sigma_{\rm kSZ}(\theta_{\rm c}) = \left(\frac{T_0\tau(\theta_{\rm c})}{c}\right)\sigma_{\rm v}\,,\,\,\,\,\, \sigma_{\rm v} = \sqrt{\sigma_{\rm v}^{(2)}}\,,
\eeq
and the noise
\beq
\sigma_{\rm N}^2(\sigma_{\rm v}) = \sum^{\ell_{\rm max}}_\ell\frac{2\ell+1}{4\pi}\langle C^{\rm obs}_\ell\rangle U(\ell\theta_{\rm c}) U(\ell\theta_{\rm c}) \,.
\eeq
The ensemble average of the observed CMB angular power spectrum $C^{\rm obs}_\ell = B^2_\ell C^{\rm the}_\ell + N_\ell$, with a Gaussian beam function $B_\ell$,  theoretical prediction of the CMB power spectrum $C^{\rm the}_\ell$ and the detector noise $N_\ell$.

We choose the $\theta_{\rm c}$ that minimizes $R_{\rm N}$ and calculate the corresponding $\tau_T$ and ${\rm Cov}(P_{{\rm kSZ},\ell_1}(k), P_{{\rm kSZ},\ell_2}(k))$.

\bibliography{mybib}

\begin{thebibliography}{35}%
\makeatletter
\providecommand \@ifxundefined [1]{%
 \@ifx{#1\undefined}
}%
\providecommand \@ifnum [1]{%
 \ifnum #1\expandafter \@firstoftwo
 \else \expandafter \@secondoftwo
 \fi
}%
\providecommand \@ifx [1]{%
 \ifx #1\expandafter \@firstoftwo
 \else \expandafter \@secondoftwo
 \fi
}%
\providecommand \natexlab [1]{#1}%
\providecommand \enquote  [1]{``#1''}%
\providecommand \bibnamefont  [1]{#1}%
\providecommand \bibfnamefont [1]{#1}%
\providecommand \citenamefont [1]{#1}%
\providecommand \href@noop [0]{\@secondoftwo}%
\providecommand \href [0]{\begingroup \@sanitize@url \@href}%
\providecommand \@href[1]{\@@startlink{#1}\@@href}%
\providecommand \@@href[1]{\endgroup#1\@@endlink}%
\providecommand \@sanitize@url [0]{\catcode `\\12\catcode `\$12\catcode
  `\&12\catcode `\#12\catcode `\^12\catcode `\_12\catcode `\%12\relax}%
\providecommand \@@startlink[1]{}%
\providecommand \@@endlink[0]{}%
\providecommand \url  [0]{\begingroup\@sanitize@url \@url }%
\providecommand \@url [1]{\endgroup\@href {#1}{\urlprefix }}%
\providecommand \urlprefix  [0]{URL }%
\providecommand \Eprint [0]{\href }%
\providecommand \doibase [0]{http://dx.doi.org/}%
\providecommand \selectlanguage [0]{\@gobble}%
\providecommand \bibinfo  [0]{\@secondoftwo}%
\providecommand \bibfield  [0]{\@secondoftwo}%
\providecommand \translation [1]{[#1]}%
\providecommand \BibitemOpen [0]{}%
\providecommand \bibitemStop [0]{}%
\providecommand \bibitemNoStop [0]{.\EOS\space}%
\providecommand \EOS [0]{\spacefactor3000\relax}%
\providecommand \BibitemShut  [1]{\csname bibitem#1\endcsname}%
\let\auto@bib@innerbib\@empty
\bibitem [{\citenamefont {{Sunyaev}}\ and\ \citenamefont
  {{Zeldovich}}(1970)}]{kSZ1970}%
  \BibitemOpen
  \bibfield  {author} {\bibinfo {author} {\bibfnamefont {R.~A.}\ \bibnamefont
  {{Sunyaev}}}\ and\ \bibinfo {author} {\bibfnamefont {Y.~B.}\ \bibnamefont
  {{Zeldovich}}},\ }\href {\doibase 10.1007/BF00653471} {\bibfield  {journal}
  {\bibinfo  {journal} {\apss}\ }\textbf {\bibinfo {volume} {7}},\ \bibinfo
  {pages} {3} (\bibinfo {year} {1970})}\BibitemShut {NoStop}%
\bibitem [{\citenamefont {{Sunyaev}}\ and\ \citenamefont
  {{Zeldovich}}(1972)}]{kSZ1972}%
  \BibitemOpen
  \bibfield  {author} {\bibinfo {author} {\bibfnamefont {R.~A.}\ \bibnamefont
  {{Sunyaev}}}\ and\ \bibinfo {author} {\bibfnamefont {Y.~B.}\ \bibnamefont
  {{Zeldovich}}},\ }\href@noop {} {\bibfield  {journal} {\bibinfo  {journal}
  {Comments on Astrophysics and Space Physics}\ }\textbf {\bibinfo {volume}
  {4}},\ \bibinfo {pages} {173} (\bibinfo {year} {1972})}\BibitemShut {NoStop}%
\bibitem [{\citenamefont {{Sunyaev}}\ and\ \citenamefont
  {{Zeldovich}}(1980)}]{kSZ1980}%
  \BibitemOpen
  \bibfield  {author} {\bibinfo {author} {\bibfnamefont {R.~A.}\ \bibnamefont
  {{Sunyaev}}}\ and\ \bibinfo {author} {\bibfnamefont {I.~B.}\ \bibnamefont
  {{Zeldovich}}},\ }\href {\doibase 10.1093/mnras/190.3.413} {\bibfield
  {journal} {\bibinfo  {journal} {\mnras}\ }\textbf {\bibinfo {volume} {190}},\
  \bibinfo {pages} {413} (\bibinfo {year} {1980})}\BibitemShut {NoStop}%
\bibitem [{\citenamefont {{Ostriker}}\ and\ \citenamefont
  {{Vishniac}}(1986)}]{kSZ1986}%
  \BibitemOpen
  \bibfield  {author} {\bibinfo {author} {\bibfnamefont {J.~P.}\ \bibnamefont
  {{Ostriker}}}\ and\ \bibinfo {author} {\bibfnamefont {E.~T.}\ \bibnamefont
  {{Vishniac}}},\ }\href {\doibase 10.1086/184704} {\bibfield  {journal}
  {\bibinfo  {journal} {\apjl}\ }\textbf {\bibinfo {volume} {306}},\ \bibinfo
  {pages} {L51} (\bibinfo {year} {1986})}\BibitemShut {NoStop}%
\bibitem [{\citenamefont {{Hand}}\ \emph {et~al.}(2012)\citenamefont {{Hand}},
  \citenamefont {{Addison}}, \citenamefont {{Aubourg}}, \citenamefont
  {{Battaglia}}, \citenamefont {{Battistelli}}, \citenamefont {{Bizyaev}},
  \citenamefont {{Bond}}, \citenamefont {{Brewington}}, \citenamefont
  {{Brinkmann}}, \citenamefont {{Brown}}, \citenamefont {{Das}}, \citenamefont
  {{Dawson}}, \citenamefont {{Devlin}}, \citenamefont {{Dunkley}},
  \citenamefont {{Dunner}}, \citenamefont {{Eisenstein}}, \citenamefont
  {{Fowler}}, \citenamefont {{Gralla}}, \citenamefont {{Hajian}}, \citenamefont
  {{Halpern}}, \citenamefont {{Hilton}}, \citenamefont {{Hincks}},
  \citenamefont {{Hlozek}}, \citenamefont {{Hughes}}, \citenamefont
  {{Infante}}, \citenamefont {{Irwin}}, \citenamefont {{Kosowsky}},
  \citenamefont {{Lin}}, \citenamefont {{Malanushenko}}, \citenamefont
  {{Malanushenko}}, \citenamefont {{Marriage}}, \citenamefont {{Marsden}},
  \citenamefont {{Menanteau}}, \citenamefont {{Moodley}}, \citenamefont
  {{Niemack}}, \citenamefont {{Nolta}}, \citenamefont {{Oravetz}},
  \citenamefont {{Page}}, \citenamefont {{Palanque-Delabrouille}},
  \citenamefont {{Pan}}, \citenamefont {{Reese}}, \citenamefont {{Schlegel}},
  \citenamefont {{Schneider}}, \citenamefont {{Sehgal}}, \citenamefont
  {{Shelden}}, \citenamefont {{Sievers}}, \citenamefont {{Sif{\'o}n}},
  \citenamefont {{Simmons}}, \citenamefont {{Snedden}}, \citenamefont
  {{Spergel}}, \citenamefont {{Staggs}}, \citenamefont {{Swetz}}, \citenamefont
  {{Switzer}}, \citenamefont {{Trac}}, \citenamefont {{Weaver}}, \citenamefont
  {{Wollack}}, \citenamefont {{Yeche}},\ and\ \citenamefont
  {{Zunckel}}}]{Hand12}%
  \BibitemOpen
  \bibfield  {author} {\bibinfo {author} {\bibfnamefont {N.}~\bibnamefont
  {{Hand}}}, \bibinfo {author} {\bibfnamefont {G.~E.}\ \bibnamefont
  {{Addison}}}, \bibinfo {author} {\bibfnamefont {E.}~\bibnamefont
  {{Aubourg}}}, \bibinfo {author} {\bibfnamefont {N.}~\bibnamefont
  {{Battaglia}}}, \bibinfo {author} {\bibfnamefont {E.~S.}\ \bibnamefont
  {{Battistelli}}}, \bibinfo {author} {\bibfnamefont {D.}~\bibnamefont
  {{Bizyaev}}}, \bibinfo {author} {\bibfnamefont {J.~R.}\ \bibnamefont
  {{Bond}}}, \bibinfo {author} {\bibfnamefont {H.}~\bibnamefont
  {{Brewington}}}, \bibinfo {author} {\bibfnamefont {J.}~\bibnamefont
  {{Brinkmann}}}, \bibinfo {author} {\bibfnamefont {B.~R.}\ \bibnamefont
  {{Brown}}}, \bibinfo {author} {\bibfnamefont {S.}~\bibnamefont {{Das}}},
  \bibinfo {author} {\bibfnamefont {K.~S.}\ \bibnamefont {{Dawson}}}, \bibinfo
  {author} {\bibfnamefont {M.~J.}\ \bibnamefont {{Devlin}}}, \bibinfo {author}
  {\bibfnamefont {J.}~\bibnamefont {{Dunkley}}}, \bibinfo {author}
  {\bibfnamefont {R.}~\bibnamefont {{Dunner}}}, \bibinfo {author}
  {\bibfnamefont {D.~J.}\ \bibnamefont {{Eisenstein}}}, \bibinfo {author}
  {\bibfnamefont {J.~W.}\ \bibnamefont {{Fowler}}}, \bibinfo {author}
  {\bibfnamefont {M.~B.}\ \bibnamefont {{Gralla}}}, \bibinfo {author}
  {\bibfnamefont {A.}~\bibnamefont {{Hajian}}}, \bibinfo {author}
  {\bibfnamefont {M.}~\bibnamefont {{Halpern}}}, \bibinfo {author}
  {\bibfnamefont {M.}~\bibnamefont {{Hilton}}}, \bibinfo {author}
  {\bibfnamefont {A.~D.}\ \bibnamefont {{Hincks}}}, \bibinfo {author}
  {\bibfnamefont {R.}~\bibnamefont {{Hlozek}}}, \bibinfo {author}
  {\bibfnamefont {J.~P.}\ \bibnamefont {{Hughes}}}, \bibinfo {author}
  {\bibfnamefont {L.}~\bibnamefont {{Infante}}}, \bibinfo {author}
  {\bibfnamefont {K.~D.}\ \bibnamefont {{Irwin}}}, \bibinfo {author}
  {\bibfnamefont {A.}~\bibnamefont {{Kosowsky}}}, \bibinfo {author}
  {\bibfnamefont {Y.-T.}\ \bibnamefont {{Lin}}}, \bibinfo {author}
  {\bibfnamefont {E.}~\bibnamefont {{Malanushenko}}}, \bibinfo {author}
  {\bibfnamefont {V.}~\bibnamefont {{Malanushenko}}}, \bibinfo {author}
  {\bibfnamefont {T.~A.}\ \bibnamefont {{Marriage}}}, \bibinfo {author}
  {\bibfnamefont {D.}~\bibnamefont {{Marsden}}}, \bibinfo {author}
  {\bibfnamefont {F.}~\bibnamefont {{Menanteau}}}, \bibinfo {author}
  {\bibfnamefont {K.}~\bibnamefont {{Moodley}}}, \bibinfo {author}
  {\bibfnamefont {M.~D.}\ \bibnamefont {{Niemack}}}, \bibinfo {author}
  {\bibfnamefont {M.~R.}\ \bibnamefont {{Nolta}}}, \bibinfo {author}
  {\bibfnamefont {D.}~\bibnamefont {{Oravetz}}}, \bibinfo {author}
  {\bibfnamefont {L.~A.}\ \bibnamefont {{Page}}}, \bibinfo {author}
  {\bibfnamefont {N.}~\bibnamefont {{Palanque-Delabrouille}}}, \bibinfo
  {author} {\bibfnamefont {K.}~\bibnamefont {{Pan}}}, \bibinfo {author}
  {\bibfnamefont {E.~D.}\ \bibnamefont {{Reese}}}, \bibinfo {author}
  {\bibfnamefont {D.~J.}\ \bibnamefont {{Schlegel}}}, \bibinfo {author}
  {\bibfnamefont {D.~P.}\ \bibnamefont {{Schneider}}}, \bibinfo {author}
  {\bibfnamefont {N.}~\bibnamefont {{Sehgal}}}, \bibinfo {author}
  {\bibfnamefont {A.}~\bibnamefont {{Shelden}}}, \bibinfo {author}
  {\bibfnamefont {J.}~\bibnamefont {{Sievers}}}, \bibinfo {author}
  {\bibfnamefont {C.}~\bibnamefont {{Sif{\'o}n}}}, \bibinfo {author}
  {\bibfnamefont {A.}~\bibnamefont {{Simmons}}}, \bibinfo {author}
  {\bibfnamefont {S.}~\bibnamefont {{Snedden}}}, \bibinfo {author}
  {\bibfnamefont {D.~N.}\ \bibnamefont {{Spergel}}}, \bibinfo {author}
  {\bibfnamefont {S.~T.}\ \bibnamefont {{Staggs}}}, \bibinfo {author}
  {\bibfnamefont {D.~S.}\ \bibnamefont {{Swetz}}}, \bibinfo {author}
  {\bibfnamefont {E.~R.}\ \bibnamefont {{Switzer}}}, \bibinfo {author}
  {\bibfnamefont {H.}~\bibnamefont {{Trac}}}, \bibinfo {author} {\bibfnamefont
  {B.~A.}\ \bibnamefont {{Weaver}}}, \bibinfo {author} {\bibfnamefont {E.~J.}\
  \bibnamefont {{Wollack}}}, \bibinfo {author} {\bibfnamefont {C.}~\bibnamefont
  {{Yeche}}}, \ and\ \bibinfo {author} {\bibfnamefont {C.}~\bibnamefont
  {{Zunckel}}},\ }\href {\doibase 10.1103/PhysRevLett.109.041101} {\bibfield
  {journal} {\bibinfo  {journal} {\prl}\ }\textbf {\bibinfo {volume} {109}},\
  \bibinfo {eid} {041101} (\bibinfo {year} {2012})},\ \Eprint
  {http://arxiv.org/abs/1203.4219} {arXiv:1203.4219 [astro-ph.CO]} \BibitemShut
  {NoStop}%
\bibitem [{\citenamefont {{Soergel}}\ \emph {et~al.}(2016)\citenamefont
  {{Soergel}}, \citenamefont {{Flender}}, \citenamefont {{Story}},
  \citenamefont {{Bleem}}, \citenamefont {{Giannantonio}}, \citenamefont
  {{Efstathiou}}, \citenamefont {{Rykoff}}, \citenamefont {{Benson}},
  \citenamefont {{Crawford}}, \citenamefont {{Dodelson}}, \citenamefont
  {{Habib}}, \citenamefont {{Heitmann}}, \citenamefont {{Holder}},
  \citenamefont {{Jain}}, \citenamefont {{Rozo}}, \citenamefont {{Saro}},
  \citenamefont {{Weller}}, \citenamefont {{Abdalla}}, \citenamefont {{Allam}},
  \citenamefont {{Annis}}, \citenamefont {{Armstrong}}, \citenamefont
  {{Benoit-L{\'e}vy}}, \citenamefont {{Bernstein}}, \citenamefont
  {{Carlstrom}}, \citenamefont {{Carnero Rosell}}, \citenamefont {{Carrasco
  Kind}}, \citenamefont {{Castander}}, \citenamefont {{Chiu}}, \citenamefont
  {{Chown}}, \citenamefont {{Crocce}}, \citenamefont {{Cunha}}, \citenamefont
  {{D'Andrea}}, \citenamefont {{da Costa}}, \citenamefont {{de Haan}},
  \citenamefont {{Desai}}, \citenamefont {{Diehl}}, \citenamefont {{Dietrich}},
  \citenamefont {{Doel}}, \citenamefont {{Estrada}}, \citenamefont {{Evrard}},
  \citenamefont {{Flaugher}}, \citenamefont {{Fosalba}}, \citenamefont
  {{Frieman}}, \citenamefont {{Gaztanaga}}, \citenamefont {{Gruen}},
  \citenamefont {{Gruendl}}, \citenamefont {{Holzapfel}}, \citenamefont
  {{Honscheid}}, \citenamefont {{James}}, \citenamefont {{Keisler}},
  \citenamefont {{Kuehn}}, \citenamefont {{Kuropatkin}}, \citenamefont
  {{Lahav}}, \citenamefont {{Lima}}, \citenamefont {{Marshall}}, \citenamefont
  {{McDonald}}, \citenamefont {{Melchior}}, \citenamefont {{Miller}},
  \citenamefont {{Miquel}}, \citenamefont {{Nord}}, \citenamefont {{Ogando}},
  \citenamefont {{Omori}}, \citenamefont {{Plazas}}, \citenamefont {{Rapetti}},
  \citenamefont {{Reichardt}}, \citenamefont {{Romer}}, \citenamefont
  {{Roodman}}, \citenamefont {{Saliwanchik}}, \citenamefont {{Sanchez}},
  \citenamefont {{Schubnell}}, \citenamefont {{Sevilla-Noarbe}}, \citenamefont
  {{Sheldon}}, \citenamefont {{Smith}}, \citenamefont {{Soares-Santos}},
  \citenamefont {{Sobreira}}, \citenamefont {{Stark}}, \citenamefont
  {{Suchyta}}, \citenamefont {{Swanson}}, \citenamefont {{Tarle}},
  \citenamefont {{Thomas}}, \citenamefont {{Vieira}}, \citenamefont {{Walker}},
  \citenamefont {{Whitehorn}}, \citenamefont {{DES Collaboration}},\ and\
  \citenamefont {{SPT Collaboration}}}]{Soergel16}%
  \BibitemOpen
  \bibfield  {author} {\bibinfo {author} {\bibfnamefont {B.}~\bibnamefont
  {{Soergel}}}, \bibinfo {author} {\bibfnamefont {S.}~\bibnamefont
  {{Flender}}}, \bibinfo {author} {\bibfnamefont {K.~T.}\ \bibnamefont
  {{Story}}}, \bibinfo {author} {\bibfnamefont {L.}~\bibnamefont {{Bleem}}},
  \bibinfo {author} {\bibfnamefont {T.}~\bibnamefont {{Giannantonio}}},
  \bibinfo {author} {\bibfnamefont {G.}~\bibnamefont {{Efstathiou}}}, \bibinfo
  {author} {\bibfnamefont {E.}~\bibnamefont {{Rykoff}}}, \bibinfo {author}
  {\bibfnamefont {B.~A.}\ \bibnamefont {{Benson}}}, \bibinfo {author}
  {\bibfnamefont {T.}~\bibnamefont {{Crawford}}}, \bibinfo {author}
  {\bibfnamefont {S.}~\bibnamefont {{Dodelson}}}, \bibinfo {author}
  {\bibfnamefont {S.}~\bibnamefont {{Habib}}}, \bibinfo {author} {\bibfnamefont
  {K.}~\bibnamefont {{Heitmann}}}, \bibinfo {author} {\bibfnamefont
  {G.}~\bibnamefont {{Holder}}}, \bibinfo {author} {\bibfnamefont
  {B.}~\bibnamefont {{Jain}}}, \bibinfo {author} {\bibfnamefont
  {E.}~\bibnamefont {{Rozo}}}, \bibinfo {author} {\bibfnamefont
  {A.}~\bibnamefont {{Saro}}}, \bibinfo {author} {\bibfnamefont
  {J.}~\bibnamefont {{Weller}}}, \bibinfo {author} {\bibfnamefont {F.~B.}\
  \bibnamefont {{Abdalla}}}, \bibinfo {author} {\bibfnamefont {S.}~\bibnamefont
  {{Allam}}}, \bibinfo {author} {\bibfnamefont {J.}~\bibnamefont {{Annis}}},
  \bibinfo {author} {\bibfnamefont {R.}~\bibnamefont {{Armstrong}}}, \bibinfo
  {author} {\bibfnamefont {A.}~\bibnamefont {{Benoit-L{\'e}vy}}}, \bibinfo
  {author} {\bibfnamefont {G.~M.}\ \bibnamefont {{Bernstein}}}, \bibinfo
  {author} {\bibfnamefont {J.~E.}\ \bibnamefont {{Carlstrom}}}, \bibinfo
  {author} {\bibfnamefont {A.}~\bibnamefont {{Carnero Rosell}}}, \bibinfo
  {author} {\bibfnamefont {M.}~\bibnamefont {{Carrasco Kind}}}, \bibinfo
  {author} {\bibfnamefont {F.~J.}\ \bibnamefont {{Castander}}}, \bibinfo
  {author} {\bibfnamefont {I.}~\bibnamefont {{Chiu}}}, \bibinfo {author}
  {\bibfnamefont {R.}~\bibnamefont {{Chown}}}, \bibinfo {author} {\bibfnamefont
  {M.}~\bibnamefont {{Crocce}}}, \bibinfo {author} {\bibfnamefont {C.~E.}\
  \bibnamefont {{Cunha}}}, \bibinfo {author} {\bibfnamefont {C.~B.}\
  \bibnamefont {{D'Andrea}}}, \bibinfo {author} {\bibfnamefont {L.~N.}\
  \bibnamefont {{da Costa}}}, \bibinfo {author} {\bibfnamefont
  {T.}~\bibnamefont {{de Haan}}}, \bibinfo {author} {\bibfnamefont
  {S.}~\bibnamefont {{Desai}}}, \bibinfo {author} {\bibfnamefont {H.~T.}\
  \bibnamefont {{Diehl}}}, \bibinfo {author} {\bibfnamefont {J.~P.}\
  \bibnamefont {{Dietrich}}}, \bibinfo {author} {\bibfnamefont
  {P.}~\bibnamefont {{Doel}}}, \bibinfo {author} {\bibfnamefont
  {J.}~\bibnamefont {{Estrada}}}, \bibinfo {author} {\bibfnamefont {A.~E.}\
  \bibnamefont {{Evrard}}}, \bibinfo {author} {\bibfnamefont {B.}~\bibnamefont
  {{Flaugher}}}, \bibinfo {author} {\bibfnamefont {P.}~\bibnamefont
  {{Fosalba}}}, \bibinfo {author} {\bibfnamefont {J.}~\bibnamefont
  {{Frieman}}}, \bibinfo {author} {\bibfnamefont {E.}~\bibnamefont
  {{Gaztanaga}}}, \bibinfo {author} {\bibfnamefont {D.}~\bibnamefont
  {{Gruen}}}, \bibinfo {author} {\bibfnamefont {R.~A.}\ \bibnamefont
  {{Gruendl}}}, \bibinfo {author} {\bibfnamefont {W.~L.}\ \bibnamefont
  {{Holzapfel}}}, \bibinfo {author} {\bibfnamefont {K.}~\bibnamefont
  {{Honscheid}}}, \bibinfo {author} {\bibfnamefont {D.~J.}\ \bibnamefont
  {{James}}}, \bibinfo {author} {\bibfnamefont {R.}~\bibnamefont {{Keisler}}},
  \bibinfo {author} {\bibfnamefont {K.}~\bibnamefont {{Kuehn}}}, \bibinfo
  {author} {\bibfnamefont {N.}~\bibnamefont {{Kuropatkin}}}, \bibinfo {author}
  {\bibfnamefont {O.}~\bibnamefont {{Lahav}}}, \bibinfo {author} {\bibfnamefont
  {M.}~\bibnamefont {{Lima}}}, \bibinfo {author} {\bibfnamefont {J.~L.}\
  \bibnamefont {{Marshall}}}, \bibinfo {author} {\bibfnamefont
  {M.}~\bibnamefont {{McDonald}}}, \bibinfo {author} {\bibfnamefont
  {P.}~\bibnamefont {{Melchior}}}, \bibinfo {author} {\bibfnamefont {C.~J.}\
  \bibnamefont {{Miller}}}, \bibinfo {author} {\bibfnamefont {R.}~\bibnamefont
  {{Miquel}}}, \bibinfo {author} {\bibfnamefont {B.}~\bibnamefont {{Nord}}},
  \bibinfo {author} {\bibfnamefont {R.}~\bibnamefont {{Ogando}}}, \bibinfo
  {author} {\bibfnamefont {Y.}~\bibnamefont {{Omori}}}, \bibinfo {author}
  {\bibfnamefont {A.~A.}\ \bibnamefont {{Plazas}}}, \bibinfo {author}
  {\bibfnamefont {D.}~\bibnamefont {{Rapetti}}}, \bibinfo {author}
  {\bibfnamefont {C.~L.}\ \bibnamefont {{Reichardt}}}, \bibinfo {author}
  {\bibfnamefont {A.~K.}\ \bibnamefont {{Romer}}}, \bibinfo {author}
  {\bibfnamefont {A.}~\bibnamefont {{Roodman}}}, \bibinfo {author}
  {\bibfnamefont {B.~R.}\ \bibnamefont {{Saliwanchik}}}, \bibinfo {author}
  {\bibfnamefont {E.}~\bibnamefont {{Sanchez}}}, \bibinfo {author}
  {\bibfnamefont {M.}~\bibnamefont {{Schubnell}}}, \bibinfo {author}
  {\bibfnamefont {I.}~\bibnamefont {{Sevilla-Noarbe}}}, \bibinfo {author}
  {\bibfnamefont {E.}~\bibnamefont {{Sheldon}}}, \bibinfo {author}
  {\bibfnamefont {R.~C.}\ \bibnamefont {{Smith}}}, \bibinfo {author}
  {\bibfnamefont {M.}~\bibnamefont {{Soares-Santos}}}, \bibinfo {author}
  {\bibfnamefont {F.}~\bibnamefont {{Sobreira}}}, \bibinfo {author}
  {\bibfnamefont {A.}~\bibnamefont {{Stark}}}, \bibinfo {author} {\bibfnamefont
  {E.}~\bibnamefont {{Suchyta}}}, \bibinfo {author} {\bibfnamefont {M.~E.~C.}\
  \bibnamefont {{Swanson}}}, \bibinfo {author} {\bibfnamefont {G.}~\bibnamefont
  {{Tarle}}}, \bibinfo {author} {\bibfnamefont {D.}~\bibnamefont {{Thomas}}},
  \bibinfo {author} {\bibfnamefont {J.~D.}\ \bibnamefont {{Vieira}}}, \bibinfo
  {author} {\bibfnamefont {A.~R.}\ \bibnamefont {{Walker}}}, \bibinfo {author}
  {\bibfnamefont {N.}~\bibnamefont {{Whitehorn}}}, \bibinfo {author}
  {\bibnamefont {{DES Collaboration}}}, \ and\ \bibinfo {author} {\bibnamefont
  {{SPT Collaboration}}},\ }\href {\doibase 10.1093/mnras/stw1455} {\bibfield
  {journal} {\bibinfo  {journal} {\mnras}\ }\textbf {\bibinfo {volume} {461}},\
  \bibinfo {pages} {3172} (\bibinfo {year} {2016})},\ \Eprint
  {http://arxiv.org/abs/1603.03904} {arXiv:1603.03904 [astro-ph.CO]}
  \BibitemShut {NoStop}%
\bibitem [{\citenamefont {{Schaan}}\ \emph {et~al.}(2016)\citenamefont
  {{Schaan}}, \citenamefont {{Ferraro}}, \citenamefont {{Vargas-Maga{\~n}a}},
  \citenamefont {{Smith}}, \citenamefont {{Ho}}, \citenamefont {{Aiola}},
  \citenamefont {{Battaglia}}, \citenamefont {{Bond}}, \citenamefont {{De
  Bernardis}}, \citenamefont {{Calabrese}}, \citenamefont {{Cho}},
  \citenamefont {{Devlin}}, \citenamefont {{Dunkley}}, \citenamefont
  {{Gallardo}}, \citenamefont {{Hasselfield}}, \citenamefont {{Henderson}},
  \citenamefont {{Hill}}, \citenamefont {{Hincks}}, \citenamefont {{Hlozek}},
  \citenamefont {{Hubmayr}}, \citenamefont {{Hughes}}, \citenamefont {{Irwin}},
  \citenamefont {{Koopman}}, \citenamefont {{Kosowsky}}, \citenamefont {{Li}},
  \citenamefont {{Louis}}, \citenamefont {{Lungu}}, \citenamefont
  {{Madhavacheril}}, \citenamefont {{Maurin}}, \citenamefont {{McMahon}},
  \citenamefont {{Moodley}}, \citenamefont {{Naess}}, \citenamefont {{Nati}},
  \citenamefont {{Newburgh}}, \citenamefont {{Niemack}}, \citenamefont
  {{Page}}, \citenamefont {{Pappas}}, \citenamefont {{Partridge}},
  \citenamefont {{Schmitt}}, \citenamefont {{Sehgal}}, \citenamefont
  {{Sherwin}}, \citenamefont {{Sievers}}, \citenamefont {{Spergel}},
  \citenamefont {{Staggs}}, \citenamefont {{van Engelen}}, \citenamefont
  {{Wollack}},\ and\ \citenamefont {{ACTPol Collaboration}}}]{Schaan16}%
  \BibitemOpen
  \bibfield  {author} {\bibinfo {author} {\bibfnamefont {E.}~\bibnamefont
  {{Schaan}}}, \bibinfo {author} {\bibfnamefont {S.}~\bibnamefont {{Ferraro}}},
  \bibinfo {author} {\bibfnamefont {M.}~\bibnamefont {{Vargas-Maga{\~n}a}}},
  \bibinfo {author} {\bibfnamefont {K.~M.}\ \bibnamefont {{Smith}}}, \bibinfo
  {author} {\bibfnamefont {S.}~\bibnamefont {{Ho}}}, \bibinfo {author}
  {\bibfnamefont {S.}~\bibnamefont {{Aiola}}}, \bibinfo {author} {\bibfnamefont
  {N.}~\bibnamefont {{Battaglia}}}, \bibinfo {author} {\bibfnamefont {J.~R.}\
  \bibnamefont {{Bond}}}, \bibinfo {author} {\bibfnamefont {F.}~\bibnamefont
  {{De Bernardis}}}, \bibinfo {author} {\bibfnamefont {E.}~\bibnamefont
  {{Calabrese}}}, \bibinfo {author} {\bibfnamefont {H.-M.}\ \bibnamefont
  {{Cho}}}, \bibinfo {author} {\bibfnamefont {M.~J.}\ \bibnamefont {{Devlin}}},
  \bibinfo {author} {\bibfnamefont {J.}~\bibnamefont {{Dunkley}}}, \bibinfo
  {author} {\bibfnamefont {P.~A.}\ \bibnamefont {{Gallardo}}}, \bibinfo
  {author} {\bibfnamefont {M.}~\bibnamefont {{Hasselfield}}}, \bibinfo {author}
  {\bibfnamefont {S.}~\bibnamefont {{Henderson}}}, \bibinfo {author}
  {\bibfnamefont {J.~C.}\ \bibnamefont {{Hill}}}, \bibinfo {author}
  {\bibfnamefont {A.~D.}\ \bibnamefont {{Hincks}}}, \bibinfo {author}
  {\bibfnamefont {R.}~\bibnamefont {{Hlozek}}}, \bibinfo {author}
  {\bibfnamefont {J.}~\bibnamefont {{Hubmayr}}}, \bibinfo {author}
  {\bibfnamefont {J.~P.}\ \bibnamefont {{Hughes}}}, \bibinfo {author}
  {\bibfnamefont {K.~D.}\ \bibnamefont {{Irwin}}}, \bibinfo {author}
  {\bibfnamefont {B.}~\bibnamefont {{Koopman}}}, \bibinfo {author}
  {\bibfnamefont {A.}~\bibnamefont {{Kosowsky}}}, \bibinfo {author}
  {\bibfnamefont {D.}~\bibnamefont {{Li}}}, \bibinfo {author} {\bibfnamefont
  {T.}~\bibnamefont {{Louis}}}, \bibinfo {author} {\bibfnamefont
  {M.}~\bibnamefont {{Lungu}}}, \bibinfo {author} {\bibfnamefont
  {M.}~\bibnamefont {{Madhavacheril}}}, \bibinfo {author} {\bibfnamefont
  {L.}~\bibnamefont {{Maurin}}}, \bibinfo {author} {\bibfnamefont {J.~J.}\
  \bibnamefont {{McMahon}}}, \bibinfo {author} {\bibfnamefont {K.}~\bibnamefont
  {{Moodley}}}, \bibinfo {author} {\bibfnamefont {S.}~\bibnamefont {{Naess}}},
  \bibinfo {author} {\bibfnamefont {F.}~\bibnamefont {{Nati}}}, \bibinfo
  {author} {\bibfnamefont {L.}~\bibnamefont {{Newburgh}}}, \bibinfo {author}
  {\bibfnamefont {M.~D.}\ \bibnamefont {{Niemack}}}, \bibinfo {author}
  {\bibfnamefont {L.~A.}\ \bibnamefont {{Page}}}, \bibinfo {author}
  {\bibfnamefont {C.~G.}\ \bibnamefont {{Pappas}}}, \bibinfo {author}
  {\bibfnamefont {B.}~\bibnamefont {{Partridge}}}, \bibinfo {author}
  {\bibfnamefont {B.~L.}\ \bibnamefont {{Schmitt}}}, \bibinfo {author}
  {\bibfnamefont {N.}~\bibnamefont {{Sehgal}}}, \bibinfo {author}
  {\bibfnamefont {B.~D.}\ \bibnamefont {{Sherwin}}}, \bibinfo {author}
  {\bibfnamefont {J.~L.}\ \bibnamefont {{Sievers}}}, \bibinfo {author}
  {\bibfnamefont {D.~N.}\ \bibnamefont {{Spergel}}}, \bibinfo {author}
  {\bibfnamefont {S.~T.}\ \bibnamefont {{Staggs}}}, \bibinfo {author}
  {\bibfnamefont {A.}~\bibnamefont {{van Engelen}}}, \bibinfo {author}
  {\bibfnamefont {E.~J.}\ \bibnamefont {{Wollack}}}, \ and\ \bibinfo {author}
  {\bibnamefont {{ACTPol Collaboration}}},\ }\href {\doibase
  10.1103/PhysRevD.93.082002} {\bibfield  {journal} {\bibinfo  {journal}
  {\prd}\ }\textbf {\bibinfo {volume} {93}},\ \bibinfo {eid} {082002} (\bibinfo
  {year} {2016})},\ \Eprint {http://arxiv.org/abs/1510.06442} {arXiv:1510.06442
  [astro-ph.CO]} \BibitemShut {NoStop}%
\bibitem [{\citenamefont {{Planck Collaboration}}\ \emph
  {et~al.}(2016)\citenamefont {{Planck Collaboration}}, \citenamefont {{Ade}},
  \citenamefont {{Aghanim}}, \citenamefont {{Arnaud}}, \citenamefont
  {{Ashdown}}, \citenamefont {{Aubourg}}, \citenamefont {{Aumont}},
  \citenamefont {{Baccigalupi}}, \citenamefont {{Banday}}, \citenamefont
  {{Barreiro}}, \citenamefont {{Bartolo}}, \citenamefont {{Battaner}},
  \citenamefont {{Benabed}}, \citenamefont {{Benoit-L{\'e}vy}}, \citenamefont
  {{Bersanelli}}, \citenamefont {{Bielewicz}}, \citenamefont {{Bock}},
  \citenamefont {{Bonaldi}}, \citenamefont {{Bonavera}}, \citenamefont
  {{Bond}}, \citenamefont {{Borrill}}, \citenamefont {{Bouchet}}, \citenamefont
  {{Burigana}}, \citenamefont {{Calabrese}}, \citenamefont {{Cardoso}},
  \citenamefont {{Catalano}}, \citenamefont {{Chamballu}}, \citenamefont
  {{Chiang}}, \citenamefont {{Christensen}}, \citenamefont {{Clements}},
  \citenamefont {{Colombo}}, \citenamefont {{Combet}}, \citenamefont {{Crill}},
  \citenamefont {{Curto}}, \citenamefont {{Cuttaia}}, \citenamefont {{Danese}},
  \citenamefont {{Davies}}, \citenamefont {{Davis}}, \citenamefont {{de
  Bernardis}}, \citenamefont {{de Zotti}}, \citenamefont {{Delabrouille}},
  \citenamefont {{Dickinson}}, \citenamefont {{Diego}}, \citenamefont
  {{Dolag}}, \citenamefont {{Donzelli}}, \citenamefont {{Dor{\'e}}},
  \citenamefont {{Douspis}}, \citenamefont {{Ducout}}, \citenamefont {{Dupac}},
  \citenamefont {{Efstathiou}}, \citenamefont {{Elsner}}, \citenamefont
  {{En{\ss}lin}}, \citenamefont {{Eriksen}}, \citenamefont {{Finelli}},
  \citenamefont {{Forni}}, \citenamefont {{Frailis}}, \citenamefont
  {{Fraisse}}, \citenamefont {{Franceschi}}, \citenamefont {{Frejsel}},
  \citenamefont {{Galeotta}}, \citenamefont {{Galli}}, \citenamefont {{Ganga}},
  \citenamefont {{G{\'e}nova-Santos}}, \citenamefont {{Giard}}, \citenamefont
  {{Gjerl{\o}w}}, \citenamefont {{Gonz{\'a}lez-Nuevo}}, \citenamefont
  {{G{\'o}rski}}, \citenamefont {{Gregorio}}, \citenamefont {{Gruppuso}},
  \citenamefont {{Hansen}}, \citenamefont {{Harrison}}, \citenamefont
  {{Henrot-Versill{\'e}}}, \citenamefont {{Hern{\'a}ndez-Monteagudo}},
  \citenamefont {{Herranz}}, \citenamefont {{Hildebrand t}}, \citenamefont
  {{Hivon}}, \citenamefont {{Hobson}}, \citenamefont {{Hornstrup}},
  \citenamefont {{Huffenberger}}, \citenamefont {{Hurier}}, \citenamefont
  {{Jaffe}}, \citenamefont {{Jaffe}}, \citenamefont {{Jones}}, \citenamefont
  {{Juvela}}, \citenamefont {{Keih{\"a}nen}}, \citenamefont {{Keskitalo}},
  \citenamefont {{Kitaura}}, \citenamefont {{Kneissl}}, \citenamefont
  {{Knoche}}, \citenamefont {{Kunz}}, \citenamefont {{Kurki-Suonio}},
  \citenamefont {{Lagache}}, \citenamefont {{Lamarre}}, \citenamefont
  {{Lasenby}}, \citenamefont {{Lattanzi}}, \citenamefont {{Lawrence}},
  \citenamefont {{Leonardi}}, \citenamefont {{Le{\'o}n-Tavares}}, \citenamefont
  {{Levrier}}, \citenamefont {{Liguori}}, \citenamefont {{Lilje}},
  \citenamefont {{Linden-V{\o}rnle}}, \citenamefont {{L{\'o}pez-Caniego}},
  \citenamefont {{Lubin}}, \citenamefont {{Ma}}, \citenamefont
  {{Mac{\'\i}as-P{\'e}rez}}, \citenamefont {{Maffei}}, \citenamefont {{Maino}},
  \citenamefont {{Mak}}, \citenamefont {{Mandolesi}}, \citenamefont
  {{Mangilli}}, \citenamefont {{Maris}}, \citenamefont {{Martin}},
  \citenamefont {{Mart{\'\i}nez-Gonz{\'a}lez}}, \citenamefont {{Masi}},
  \citenamefont {{Matarrese}}, \citenamefont {{McGehee}}, \citenamefont
  {{Melchiorri}}, \citenamefont {{Mennella}}, \citenamefont {{Migliaccio}},
  \citenamefont {{Miville-Desch{\^e}nes}}, \citenamefont {{Moneti}},
  \citenamefont {{Montier}}, \citenamefont {{Morgante}}, \citenamefont
  {{Mortlock}}, \citenamefont {{Munshi}}, \citenamefont {{Murphy}},
  \citenamefont {{Naselsky}}, \citenamefont {{Nati}}, \citenamefont {{Natoli}},
  \citenamefont {{Noviello}}, \citenamefont {{Novikov}}, \citenamefont
  {{Novikov}}, \citenamefont {{Oxborrow}}, \citenamefont {{Pagano}},
  \citenamefont {{Pajot}}, \citenamefont {{Paoletti}}, \citenamefont
  {{Perdereau}}, \citenamefont {{Perotto}}, \citenamefont {{Pettorino}},
  \citenamefont {{Piacentini}}, \citenamefont {{Piat}}, \citenamefont
  {{Pierpaoli}}, \citenamefont {{Pointecouteau}}, \citenamefont {{Polenta}},
  \citenamefont {{Ponthieu}}, \citenamefont {{Pratt}}, \citenamefont {{Puget}},
  \citenamefont {{Puisieux}}, \citenamefont {{Rachen}}, \citenamefont
  {{Racine}}, \citenamefont {{Reach}}, \citenamefont {{Reinecke}},
  \citenamefont {{Remazeilles}}, \citenamefont {{Renault}}, \citenamefont
  {{Renzi}}, \citenamefont {{Ristorcelli}}, \citenamefont {{Rocha}},
  \citenamefont {{Rosset}}, \citenamefont {{Rossetti}}, \citenamefont
  {{Roudier}}, \citenamefont {{Rubi{\~n}o-Mart{\'\i}n}}, \citenamefont
  {{Rusholme}}, \citenamefont {{Sandri}}, \citenamefont {{Santos}},
  \citenamefont {{Savelainen}}, \citenamefont {{Savini}}, \citenamefont
  {{Scott}}, \citenamefont {{Spencer}}, \citenamefont {{Stolyarov}},
  \citenamefont {{Sudiwala}}, \citenamefont {{Sunyaev}}, \citenamefont
  {{Sutton}}, \citenamefont {{Suur-Uski}}, \citenamefont {{Sygnet}},
  \citenamefont {{Tauber}}, \citenamefont {{Terenzi}}, \citenamefont
  {{Toffolatti}}, \citenamefont {{Tomasi}}, \citenamefont {{Tucci}},
  \citenamefont {{Valenziano}}, \citenamefont {{Valiviita}}, \citenamefont
  {{Van Tent}}, \citenamefont {{Vielva}}, \citenamefont {{Villa}},
  \citenamefont {{Wade}}, \citenamefont {{Wandelt}}, \citenamefont {{Wang}},
  \citenamefont {{Wehus}}, \citenamefont {{Yvon}}, \citenamefont {{Zacchei}},\
  and\ \citenamefont {{Zonca}}}]{PlanckkSZ16}%
  \BibitemOpen
  \bibfield  {author} {\bibinfo {author} {\bibnamefont {{Planck
  Collaboration}}}, \bibinfo {author} {\bibfnamefont {P.~A.~R.}\ \bibnamefont
  {{Ade}}}, \bibinfo {author} {\bibfnamefont {N.}~\bibnamefont {{Aghanim}}},
  \bibinfo {author} {\bibfnamefont {M.}~\bibnamefont {{Arnaud}}}, \bibinfo
  {author} {\bibfnamefont {M.}~\bibnamefont {{Ashdown}}}, \bibinfo {author}
  {\bibfnamefont {E.}~\bibnamefont {{Aubourg}}}, \bibinfo {author}
  {\bibfnamefont {J.}~\bibnamefont {{Aumont}}}, \bibinfo {author}
  {\bibfnamefont {C.}~\bibnamefont {{Baccigalupi}}}, \bibinfo {author}
  {\bibfnamefont {A.~J.}\ \bibnamefont {{Banday}}}, \bibinfo {author}
  {\bibfnamefont {R.~B.}\ \bibnamefont {{Barreiro}}}, \bibinfo {author}
  {\bibfnamefont {N.}~\bibnamefont {{Bartolo}}}, \bibinfo {author}
  {\bibfnamefont {E.}~\bibnamefont {{Battaner}}}, \bibinfo {author}
  {\bibfnamefont {K.}~\bibnamefont {{Benabed}}}, \bibinfo {author}
  {\bibfnamefont {A.}~\bibnamefont {{Benoit-L{\'e}vy}}}, \bibinfo {author}
  {\bibfnamefont {M.}~\bibnamefont {{Bersanelli}}}, \bibinfo {author}
  {\bibfnamefont {P.}~\bibnamefont {{Bielewicz}}}, \bibinfo {author}
  {\bibfnamefont {J.~J.}\ \bibnamefont {{Bock}}}, \bibinfo {author}
  {\bibfnamefont {A.}~\bibnamefont {{Bonaldi}}}, \bibinfo {author}
  {\bibfnamefont {L.}~\bibnamefont {{Bonavera}}}, \bibinfo {author}
  {\bibfnamefont {J.~R.}\ \bibnamefont {{Bond}}}, \bibinfo {author}
  {\bibfnamefont {J.}~\bibnamefont {{Borrill}}}, \bibinfo {author}
  {\bibfnamefont {F.~R.}\ \bibnamefont {{Bouchet}}}, \bibinfo {author}
  {\bibfnamefont {C.}~\bibnamefont {{Burigana}}}, \bibinfo {author}
  {\bibfnamefont {E.}~\bibnamefont {{Calabrese}}}, \bibinfo {author}
  {\bibfnamefont {J.~F.}\ \bibnamefont {{Cardoso}}}, \bibinfo {author}
  {\bibfnamefont {A.}~\bibnamefont {{Catalano}}}, \bibinfo {author}
  {\bibfnamefont {A.}~\bibnamefont {{Chamballu}}}, \bibinfo {author}
  {\bibfnamefont {H.~C.}\ \bibnamefont {{Chiang}}}, \bibinfo {author}
  {\bibfnamefont {P.~R.}\ \bibnamefont {{Christensen}}}, \bibinfo {author}
  {\bibfnamefont {D.~L.}\ \bibnamefont {{Clements}}}, \bibinfo {author}
  {\bibfnamefont {L.~P.~L.}\ \bibnamefont {{Colombo}}}, \bibinfo {author}
  {\bibfnamefont {C.}~\bibnamefont {{Combet}}}, \bibinfo {author}
  {\bibfnamefont {B.~P.}\ \bibnamefont {{Crill}}}, \bibinfo {author}
  {\bibfnamefont {A.}~\bibnamefont {{Curto}}}, \bibinfo {author} {\bibfnamefont
  {F.}~\bibnamefont {{Cuttaia}}}, \bibinfo {author} {\bibfnamefont
  {L.}~\bibnamefont {{Danese}}}, \bibinfo {author} {\bibfnamefont {R.~D.}\
  \bibnamefont {{Davies}}}, \bibinfo {author} {\bibfnamefont {R.~J.}\
  \bibnamefont {{Davis}}}, \bibinfo {author} {\bibfnamefont {P.}~\bibnamefont
  {{de Bernardis}}}, \bibinfo {author} {\bibfnamefont {G.}~\bibnamefont {{de
  Zotti}}}, \bibinfo {author} {\bibfnamefont {J.}~\bibnamefont
  {{Delabrouille}}}, \bibinfo {author} {\bibfnamefont {C.}~\bibnamefont
  {{Dickinson}}}, \bibinfo {author} {\bibfnamefont {J.~M.}\ \bibnamefont
  {{Diego}}}, \bibinfo {author} {\bibfnamefont {K.}~\bibnamefont {{Dolag}}},
  \bibinfo {author} {\bibfnamefont {S.}~\bibnamefont {{Donzelli}}}, \bibinfo
  {author} {\bibfnamefont {O.}~\bibnamefont {{Dor{\'e}}}}, \bibinfo {author}
  {\bibfnamefont {M.}~\bibnamefont {{Douspis}}}, \bibinfo {author}
  {\bibfnamefont {A.}~\bibnamefont {{Ducout}}}, \bibinfo {author}
  {\bibfnamefont {X.}~\bibnamefont {{Dupac}}}, \bibinfo {author} {\bibfnamefont
  {G.}~\bibnamefont {{Efstathiou}}}, \bibinfo {author} {\bibfnamefont
  {F.}~\bibnamefont {{Elsner}}}, \bibinfo {author} {\bibfnamefont {T.~A.}\
  \bibnamefont {{En{\ss}lin}}}, \bibinfo {author} {\bibfnamefont {H.~K.}\
  \bibnamefont {{Eriksen}}}, \bibinfo {author} {\bibfnamefont {F.}~\bibnamefont
  {{Finelli}}}, \bibinfo {author} {\bibfnamefont {O.}~\bibnamefont {{Forni}}},
  \bibinfo {author} {\bibfnamefont {M.}~\bibnamefont {{Frailis}}}, \bibinfo
  {author} {\bibfnamefont {A.~A.}\ \bibnamefont {{Fraisse}}}, \bibinfo {author}
  {\bibfnamefont {E.}~\bibnamefont {{Franceschi}}}, \bibinfo {author}
  {\bibfnamefont {A.}~\bibnamefont {{Frejsel}}}, \bibinfo {author}
  {\bibfnamefont {S.}~\bibnamefont {{Galeotta}}}, \bibinfo {author}
  {\bibfnamefont {S.}~\bibnamefont {{Galli}}}, \bibinfo {author} {\bibfnamefont
  {K.}~\bibnamefont {{Ganga}}}, \bibinfo {author} {\bibfnamefont {R.~T.}\
  \bibnamefont {{G{\'e}nova-Santos}}}, \bibinfo {author} {\bibfnamefont
  {M.}~\bibnamefont {{Giard}}}, \bibinfo {author} {\bibfnamefont
  {E.}~\bibnamefont {{Gjerl{\o}w}}}, \bibinfo {author} {\bibfnamefont
  {J.}~\bibnamefont {{Gonz{\'a}lez-Nuevo}}}, \bibinfo {author} {\bibfnamefont
  {K.~M.}\ \bibnamefont {{G{\'o}rski}}}, \bibinfo {author} {\bibfnamefont
  {A.}~\bibnamefont {{Gregorio}}}, \bibinfo {author} {\bibfnamefont
  {A.}~\bibnamefont {{Gruppuso}}}, \bibinfo {author} {\bibfnamefont {F.~K.}\
  \bibnamefont {{Hansen}}}, \bibinfo {author} {\bibfnamefont {D.~L.}\
  \bibnamefont {{Harrison}}}, \bibinfo {author} {\bibfnamefont
  {S.}~\bibnamefont {{Henrot-Versill{\'e}}}}, \bibinfo {author} {\bibfnamefont
  {C.}~\bibnamefont {{Hern{\'a}ndez-Monteagudo}}}, \bibinfo {author}
  {\bibfnamefont {D.}~\bibnamefont {{Herranz}}}, \bibinfo {author}
  {\bibfnamefont {S.~R.}\ \bibnamefont {{Hildebrand t}}}, \bibinfo {author}
  {\bibfnamefont {E.}~\bibnamefont {{Hivon}}}, \bibinfo {author} {\bibfnamefont
  {M.}~\bibnamefont {{Hobson}}}, \bibinfo {author} {\bibfnamefont
  {A.}~\bibnamefont {{Hornstrup}}}, \bibinfo {author} {\bibfnamefont {K.~M.}\
  \bibnamefont {{Huffenberger}}}, \bibinfo {author} {\bibfnamefont
  {G.}~\bibnamefont {{Hurier}}}, \bibinfo {author} {\bibfnamefont {A.~H.}\
  \bibnamefont {{Jaffe}}}, \bibinfo {author} {\bibfnamefont {T.~R.}\
  \bibnamefont {{Jaffe}}}, \bibinfo {author} {\bibfnamefont {W.~C.}\
  \bibnamefont {{Jones}}}, \bibinfo {author} {\bibfnamefont {M.}~\bibnamefont
  {{Juvela}}}, \bibinfo {author} {\bibfnamefont {E.}~\bibnamefont
  {{Keih{\"a}nen}}}, \bibinfo {author} {\bibfnamefont {R.}~\bibnamefont
  {{Keskitalo}}}, \bibinfo {author} {\bibfnamefont {F.}~\bibnamefont
  {{Kitaura}}}, \bibinfo {author} {\bibfnamefont {R.}~\bibnamefont
  {{Kneissl}}}, \bibinfo {author} {\bibfnamefont {J.}~\bibnamefont {{Knoche}}},
  \bibinfo {author} {\bibfnamefont {M.}~\bibnamefont {{Kunz}}}, \bibinfo
  {author} {\bibfnamefont {H.}~\bibnamefont {{Kurki-Suonio}}}, \bibinfo
  {author} {\bibfnamefont {G.}~\bibnamefont {{Lagache}}}, \bibinfo {author}
  {\bibfnamefont {J.~M.}\ \bibnamefont {{Lamarre}}}, \bibinfo {author}
  {\bibfnamefont {A.}~\bibnamefont {{Lasenby}}}, \bibinfo {author}
  {\bibfnamefont {M.}~\bibnamefont {{Lattanzi}}}, \bibinfo {author}
  {\bibfnamefont {C.~R.}\ \bibnamefont {{Lawrence}}}, \bibinfo {author}
  {\bibfnamefont {R.}~\bibnamefont {{Leonardi}}}, \bibinfo {author}
  {\bibfnamefont {J.}~\bibnamefont {{Le{\'o}n-Tavares}}}, \bibinfo {author}
  {\bibfnamefont {F.}~\bibnamefont {{Levrier}}}, \bibinfo {author}
  {\bibfnamefont {M.}~\bibnamefont {{Liguori}}}, \bibinfo {author}
  {\bibfnamefont {P.~B.}\ \bibnamefont {{Lilje}}}, \bibinfo {author}
  {\bibfnamefont {M.}~\bibnamefont {{Linden-V{\o}rnle}}}, \bibinfo {author}
  {\bibfnamefont {M.}~\bibnamefont {{L{\'o}pez-Caniego}}}, \bibinfo {author}
  {\bibfnamefont {P.~M.}\ \bibnamefont {{Lubin}}}, \bibinfo {author}
  {\bibfnamefont {Y.~Z.}\ \bibnamefont {{Ma}}}, \bibinfo {author}
  {\bibfnamefont {J.~F.}\ \bibnamefont {{Mac{\'\i}as-P{\'e}rez}}}, \bibinfo
  {author} {\bibfnamefont {B.}~\bibnamefont {{Maffei}}}, \bibinfo {author}
  {\bibfnamefont {D.}~\bibnamefont {{Maino}}}, \bibinfo {author} {\bibfnamefont
  {D.~S.~Y.}\ \bibnamefont {{Mak}}}, \bibinfo {author} {\bibfnamefont
  {N.}~\bibnamefont {{Mandolesi}}}, \bibinfo {author} {\bibfnamefont
  {A.}~\bibnamefont {{Mangilli}}}, \bibinfo {author} {\bibfnamefont
  {M.}~\bibnamefont {{Maris}}}, \bibinfo {author} {\bibfnamefont {P.~G.}\
  \bibnamefont {{Martin}}}, \bibinfo {author} {\bibfnamefont {E.}~\bibnamefont
  {{Mart{\'\i}nez-Gonz{\'a}lez}}}, \bibinfo {author} {\bibfnamefont
  {S.}~\bibnamefont {{Masi}}}, \bibinfo {author} {\bibfnamefont
  {S.}~\bibnamefont {{Matarrese}}}, \bibinfo {author} {\bibfnamefont
  {P.}~\bibnamefont {{McGehee}}}, \bibinfo {author} {\bibfnamefont
  {A.}~\bibnamefont {{Melchiorri}}}, \bibinfo {author} {\bibfnamefont
  {A.}~\bibnamefont {{Mennella}}}, \bibinfo {author} {\bibfnamefont
  {M.}~\bibnamefont {{Migliaccio}}}, \bibinfo {author} {\bibfnamefont {M.~A.}\
  \bibnamefont {{Miville-Desch{\^e}nes}}}, \bibinfo {author} {\bibfnamefont
  {A.}~\bibnamefont {{Moneti}}}, \bibinfo {author} {\bibfnamefont
  {L.}~\bibnamefont {{Montier}}}, \bibinfo {author} {\bibfnamefont
  {G.}~\bibnamefont {{Morgante}}}, \bibinfo {author} {\bibfnamefont
  {D.}~\bibnamefont {{Mortlock}}}, \bibinfo {author} {\bibfnamefont
  {D.}~\bibnamefont {{Munshi}}}, \bibinfo {author} {\bibfnamefont {J.~A.}\
  \bibnamefont {{Murphy}}}, \bibinfo {author} {\bibfnamefont {P.}~\bibnamefont
  {{Naselsky}}}, \bibinfo {author} {\bibfnamefont {F.}~\bibnamefont {{Nati}}},
  \bibinfo {author} {\bibfnamefont {P.}~\bibnamefont {{Natoli}}}, \bibinfo
  {author} {\bibfnamefont {F.}~\bibnamefont {{Noviello}}}, \bibinfo {author}
  {\bibfnamefont {D.}~\bibnamefont {{Novikov}}}, \bibinfo {author}
  {\bibfnamefont {I.}~\bibnamefont {{Novikov}}}, \bibinfo {author}
  {\bibfnamefont {C.~A.}\ \bibnamefont {{Oxborrow}}}, \bibinfo {author}
  {\bibfnamefont {L.}~\bibnamefont {{Pagano}}}, \bibinfo {author}
  {\bibfnamefont {F.}~\bibnamefont {{Pajot}}}, \bibinfo {author} {\bibfnamefont
  {D.}~\bibnamefont {{Paoletti}}}, \bibinfo {author} {\bibfnamefont
  {O.}~\bibnamefont {{Perdereau}}}, \bibinfo {author} {\bibfnamefont
  {L.}~\bibnamefont {{Perotto}}}, \bibinfo {author} {\bibfnamefont
  {V.}~\bibnamefont {{Pettorino}}}, \bibinfo {author} {\bibfnamefont
  {F.}~\bibnamefont {{Piacentini}}}, \bibinfo {author} {\bibfnamefont
  {M.}~\bibnamefont {{Piat}}}, \bibinfo {author} {\bibfnamefont
  {E.}~\bibnamefont {{Pierpaoli}}}, \bibinfo {author} {\bibfnamefont
  {E.}~\bibnamefont {{Pointecouteau}}}, \bibinfo {author} {\bibfnamefont
  {G.}~\bibnamefont {{Polenta}}}, \bibinfo {author} {\bibfnamefont
  {N.}~\bibnamefont {{Ponthieu}}}, \bibinfo {author} {\bibfnamefont {G.~W.}\
  \bibnamefont {{Pratt}}}, \bibinfo {author} {\bibfnamefont {J.~L.}\
  \bibnamefont {{Puget}}}, \bibinfo {author} {\bibfnamefont {S.}~\bibnamefont
  {{Puisieux}}}, \bibinfo {author} {\bibfnamefont {J.~P.}\ \bibnamefont
  {{Rachen}}}, \bibinfo {author} {\bibfnamefont {B.}~\bibnamefont {{Racine}}},
  \bibinfo {author} {\bibfnamefont {W.~T.}\ \bibnamefont {{Reach}}}, \bibinfo
  {author} {\bibfnamefont {M.}~\bibnamefont {{Reinecke}}}, \bibinfo {author}
  {\bibfnamefont {M.}~\bibnamefont {{Remazeilles}}}, \bibinfo {author}
  {\bibfnamefont {C.}~\bibnamefont {{Renault}}}, \bibinfo {author}
  {\bibfnamefont {A.}~\bibnamefont {{Renzi}}}, \bibinfo {author} {\bibfnamefont
  {I.}~\bibnamefont {{Ristorcelli}}}, \bibinfo {author} {\bibfnamefont
  {G.}~\bibnamefont {{Rocha}}}, \bibinfo {author} {\bibfnamefont
  {C.}~\bibnamefont {{Rosset}}}, \bibinfo {author} {\bibfnamefont
  {M.}~\bibnamefont {{Rossetti}}}, \bibinfo {author} {\bibfnamefont
  {G.}~\bibnamefont {{Roudier}}}, \bibinfo {author} {\bibfnamefont {J.~A.}\
  \bibnamefont {{Rubi{\~n}o-Mart{\'\i}n}}}, \bibinfo {author} {\bibfnamefont
  {B.}~\bibnamefont {{Rusholme}}}, \bibinfo {author} {\bibfnamefont
  {M.}~\bibnamefont {{Sandri}}}, \bibinfo {author} {\bibfnamefont
  {D.}~\bibnamefont {{Santos}}}, \bibinfo {author} {\bibfnamefont
  {M.}~\bibnamefont {{Savelainen}}}, \bibinfo {author} {\bibfnamefont
  {G.}~\bibnamefont {{Savini}}}, \bibinfo {author} {\bibfnamefont
  {D.}~\bibnamefont {{Scott}}}, \bibinfo {author} {\bibfnamefont {L.~D.}\
  \bibnamefont {{Spencer}}}, \bibinfo {author} {\bibfnamefont {V.}~\bibnamefont
  {{Stolyarov}}}, \bibinfo {author} {\bibfnamefont {R.}~\bibnamefont
  {{Sudiwala}}}, \bibinfo {author} {\bibfnamefont {R.}~\bibnamefont
  {{Sunyaev}}}, \bibinfo {author} {\bibfnamefont {D.}~\bibnamefont {{Sutton}}},
  \bibinfo {author} {\bibfnamefont {A.~S.}\ \bibnamefont {{Suur-Uski}}},
  \bibinfo {author} {\bibfnamefont {J.~F.}\ \bibnamefont {{Sygnet}}}, \bibinfo
  {author} {\bibfnamefont {J.~A.}\ \bibnamefont {{Tauber}}}, \bibinfo {author}
  {\bibfnamefont {L.}~\bibnamefont {{Terenzi}}}, \bibinfo {author}
  {\bibfnamefont {L.}~\bibnamefont {{Toffolatti}}}, \bibinfo {author}
  {\bibfnamefont {M.}~\bibnamefont {{Tomasi}}}, \bibinfo {author}
  {\bibfnamefont {M.}~\bibnamefont {{Tucci}}}, \bibinfo {author} {\bibfnamefont
  {L.}~\bibnamefont {{Valenziano}}}, \bibinfo {author} {\bibfnamefont
  {J.}~\bibnamefont {{Valiviita}}}, \bibinfo {author} {\bibfnamefont
  {B.}~\bibnamefont {{Van Tent}}}, \bibinfo {author} {\bibfnamefont
  {P.}~\bibnamefont {{Vielva}}}, \bibinfo {author} {\bibfnamefont
  {F.}~\bibnamefont {{Villa}}}, \bibinfo {author} {\bibfnamefont {L.~A.}\
  \bibnamefont {{Wade}}}, \bibinfo {author} {\bibfnamefont {B.~D.}\
  \bibnamefont {{Wandelt}}}, \bibinfo {author} {\bibfnamefont {W.}~\bibnamefont
  {{Wang}}}, \bibinfo {author} {\bibfnamefont {I.~K.}\ \bibnamefont {{Wehus}}},
  \bibinfo {author} {\bibfnamefont {D.}~\bibnamefont {{Yvon}}}, \bibinfo
  {author} {\bibfnamefont {A.}~\bibnamefont {{Zacchei}}}, \ and\ \bibinfo
  {author} {\bibfnamefont {A.}~\bibnamefont {{Zonca}}},\ }\href {\doibase
  10.1051/0004-6361/201526328} {\bibfield  {journal} {\bibinfo  {journal}
  {\aap}\ }\textbf {\bibinfo {volume} {586}},\ \bibinfo {eid} {A140} (\bibinfo
  {year} {2016})},\ \Eprint {http://arxiv.org/abs/1504.03339} {arXiv:1504.03339
  [astro-ph.CO]} \BibitemShut {NoStop}%
\bibitem [{\citenamefont {{Hill}}\ \emph {et~al.}(2016)\citenamefont {{Hill}},
  \citenamefont {{Ferraro}}, \citenamefont {{Battaglia}}, \citenamefont
  {{Liu}},\ and\ \citenamefont {{Spergel}}}]{Hill16}%
  \BibitemOpen
  \bibfield  {author} {\bibinfo {author} {\bibfnamefont {J.~C.}\ \bibnamefont
  {{Hill}}}, \bibinfo {author} {\bibfnamefont {S.}~\bibnamefont {{Ferraro}}},
  \bibinfo {author} {\bibfnamefont {N.}~\bibnamefont {{Battaglia}}}, \bibinfo
  {author} {\bibfnamefont {J.}~\bibnamefont {{Liu}}}, \ and\ \bibinfo {author}
  {\bibfnamefont {D.~N.}\ \bibnamefont {{Spergel}}},\ }\href {\doibase
  10.1103/PhysRevLett.117.051301} {\bibfield  {journal} {\bibinfo  {journal}
  {\prl}\ }\textbf {\bibinfo {volume} {117}},\ \bibinfo {eid} {051301}
  (\bibinfo {year} {2016})},\ \Eprint {http://arxiv.org/abs/1603.01608}
  {arXiv:1603.01608 [astro-ph.CO]} \BibitemShut {NoStop}%
\bibitem [{\citenamefont {{De Bernardis}}\ \emph {et~al.}(2017)\citenamefont
  {{De Bernardis}}, \citenamefont {{Aiola}}, \citenamefont {{Vavagiakis}},
  \citenamefont {{Battaglia}}, \citenamefont {{Niemack}}, \citenamefont
  {{Beall}}, \citenamefont {{Becker}}, \citenamefont {{Bond}}, \citenamefont
  {{Calabrese}}, \citenamefont {{Cho}}, \citenamefont {{Coughlin}},
  \citenamefont {{Datta}}, \citenamefont {{Devlin}}, \citenamefont {{Dunkley}},
  \citenamefont {{Dunner}}, \citenamefont {{Ferraro}}, \citenamefont {{Fox}},
  \citenamefont {{Gallardo}}, \citenamefont {{Halpern}}, \citenamefont
  {{Hand}}, \citenamefont {{Hasselfield}}, \citenamefont {{Henderson}},
  \citenamefont {{Hill}}, \citenamefont {{Hilton}}, \citenamefont {{Hilton}},
  \citenamefont {{Hincks}}, \citenamefont {{Hlozek}}, \citenamefont
  {{Hubmayr}}, \citenamefont {{Huffenberger}}, \citenamefont {{Hughes}},
  \citenamefont {{Irwin}}, \citenamefont {{Koopman}}, \citenamefont
  {{Kosowsky}}, \citenamefont {{Li}}, \citenamefont {{Louis}}, \citenamefont
  {{Lungu}}, \citenamefont {{Madhavacheril}}, \citenamefont {{Maurin}},
  \citenamefont {{McMahon}}, \citenamefont {{Moodley}}, \citenamefont
  {{Naess}}, \citenamefont {{Nati}}, \citenamefont {{Newburgh}}, \citenamefont
  {{Nibarger}}, \citenamefont {{Page}}, \citenamefont {{Partridge}},
  \citenamefont {{Schaan}}, \citenamefont {{Schmitt}}, \citenamefont
  {{Sehgal}}, \citenamefont {{Sievers}}, \citenamefont {{Simon}}, \citenamefont
  {{Spergel}}, \citenamefont {{Staggs}}, \citenamefont {{Stevens}},
  \citenamefont {{Thornton}}, \citenamefont {{van Engelen}}, \citenamefont
  {{Van Lanen}},\ and\ \citenamefont {{Wollack}}}]{DeBernardis17}%
  \BibitemOpen
  \bibfield  {author} {\bibinfo {author} {\bibfnamefont {F.}~\bibnamefont {{De
  Bernardis}}}, \bibinfo {author} {\bibfnamefont {S.}~\bibnamefont {{Aiola}}},
  \bibinfo {author} {\bibfnamefont {E.~M.}\ \bibnamefont {{Vavagiakis}}},
  \bibinfo {author} {\bibfnamefont {N.}~\bibnamefont {{Battaglia}}}, \bibinfo
  {author} {\bibfnamefont {M.~D.}\ \bibnamefont {{Niemack}}}, \bibinfo {author}
  {\bibfnamefont {J.}~\bibnamefont {{Beall}}}, \bibinfo {author} {\bibfnamefont
  {D.~T.}\ \bibnamefont {{Becker}}}, \bibinfo {author} {\bibfnamefont {J.~R.}\
  \bibnamefont {{Bond}}}, \bibinfo {author} {\bibfnamefont {E.}~\bibnamefont
  {{Calabrese}}}, \bibinfo {author} {\bibfnamefont {H.}~\bibnamefont {{Cho}}},
  \bibinfo {author} {\bibfnamefont {K.}~\bibnamefont {{Coughlin}}}, \bibinfo
  {author} {\bibfnamefont {R.}~\bibnamefont {{Datta}}}, \bibinfo {author}
  {\bibfnamefont {M.}~\bibnamefont {{Devlin}}}, \bibinfo {author}
  {\bibfnamefont {J.}~\bibnamefont {{Dunkley}}}, \bibinfo {author}
  {\bibfnamefont {R.}~\bibnamefont {{Dunner}}}, \bibinfo {author}
  {\bibfnamefont {S.}~\bibnamefont {{Ferraro}}}, \bibinfo {author}
  {\bibfnamefont {A.}~\bibnamefont {{Fox}}}, \bibinfo {author} {\bibfnamefont
  {P.~A.}\ \bibnamefont {{Gallardo}}}, \bibinfo {author} {\bibfnamefont
  {M.}~\bibnamefont {{Halpern}}}, \bibinfo {author} {\bibfnamefont
  {N.}~\bibnamefont {{Hand}}}, \bibinfo {author} {\bibfnamefont
  {M.}~\bibnamefont {{Hasselfield}}}, \bibinfo {author} {\bibfnamefont {S.~W.}\
  \bibnamefont {{Henderson}}}, \bibinfo {author} {\bibfnamefont {J.~C.}\
  \bibnamefont {{Hill}}}, \bibinfo {author} {\bibfnamefont {G.~C.}\
  \bibnamefont {{Hilton}}}, \bibinfo {author} {\bibfnamefont {M.}~\bibnamefont
  {{Hilton}}}, \bibinfo {author} {\bibfnamefont {A.~D.}\ \bibnamefont
  {{Hincks}}}, \bibinfo {author} {\bibfnamefont {R.}~\bibnamefont {{Hlozek}}},
  \bibinfo {author} {\bibfnamefont {J.}~\bibnamefont {{Hubmayr}}}, \bibinfo
  {author} {\bibfnamefont {K.}~\bibnamefont {{Huffenberger}}}, \bibinfo
  {author} {\bibfnamefont {J.~P.}\ \bibnamefont {{Hughes}}}, \bibinfo {author}
  {\bibfnamefont {K.~D.}\ \bibnamefont {{Irwin}}}, \bibinfo {author}
  {\bibfnamefont {B.~J.}\ \bibnamefont {{Koopman}}}, \bibinfo {author}
  {\bibfnamefont {A.}~\bibnamefont {{Kosowsky}}}, \bibinfo {author}
  {\bibfnamefont {D.}~\bibnamefont {{Li}}}, \bibinfo {author} {\bibfnamefont
  {T.}~\bibnamefont {{Louis}}}, \bibinfo {author} {\bibfnamefont
  {M.}~\bibnamefont {{Lungu}}}, \bibinfo {author} {\bibfnamefont {M.~S.}\
  \bibnamefont {{Madhavacheril}}}, \bibinfo {author} {\bibfnamefont
  {L.}~\bibnamefont {{Maurin}}}, \bibinfo {author} {\bibfnamefont
  {J.}~\bibnamefont {{McMahon}}}, \bibinfo {author} {\bibfnamefont
  {K.}~\bibnamefont {{Moodley}}}, \bibinfo {author} {\bibfnamefont
  {S.}~\bibnamefont {{Naess}}}, \bibinfo {author} {\bibfnamefont
  {F.}~\bibnamefont {{Nati}}}, \bibinfo {author} {\bibfnamefont
  {L.}~\bibnamefont {{Newburgh}}}, \bibinfo {author} {\bibfnamefont {J.~P.}\
  \bibnamefont {{Nibarger}}}, \bibinfo {author} {\bibfnamefont {L.~A.}\
  \bibnamefont {{Page}}}, \bibinfo {author} {\bibfnamefont {B.}~\bibnamefont
  {{Partridge}}}, \bibinfo {author} {\bibfnamefont {E.}~\bibnamefont
  {{Schaan}}}, \bibinfo {author} {\bibfnamefont {B.~L.}\ \bibnamefont
  {{Schmitt}}}, \bibinfo {author} {\bibfnamefont {N.}~\bibnamefont {{Sehgal}}},
  \bibinfo {author} {\bibfnamefont {J.}~\bibnamefont {{Sievers}}}, \bibinfo
  {author} {\bibfnamefont {S.~M.}\ \bibnamefont {{Simon}}}, \bibinfo {author}
  {\bibfnamefont {D.~N.}\ \bibnamefont {{Spergel}}}, \bibinfo {author}
  {\bibfnamefont {S.~T.}\ \bibnamefont {{Staggs}}}, \bibinfo {author}
  {\bibfnamefont {J.~R.}\ \bibnamefont {{Stevens}}}, \bibinfo {author}
  {\bibfnamefont {R.~J.}\ \bibnamefont {{Thornton}}}, \bibinfo {author}
  {\bibfnamefont {A.}~\bibnamefont {{van Engelen}}}, \bibinfo {author}
  {\bibfnamefont {J.}~\bibnamefont {{Van Lanen}}}, \ and\ \bibinfo {author}
  {\bibfnamefont {E.~J.}\ \bibnamefont {{Wollack}}},\ }\href {\doibase
  10.1088/1475-7516/2017/03/008} {\bibfield  {journal} {\bibinfo  {journal}
  {\jcap}\ }\textbf {\bibinfo {volume} {2017}},\ \bibinfo {eid} {008} (\bibinfo
  {year} {2017})},\ \Eprint {http://arxiv.org/abs/1607.02139} {arXiv:1607.02139
  [astro-ph.CO]} \BibitemShut {NoStop}%
\bibitem [{\citenamefont {{Sugiyama}}\ \emph {et~al.}(2018)\citenamefont
  {{Sugiyama}}, \citenamefont {{Okumura}},\ and\ \citenamefont
  {{Spergel}}}]{Sugiyama18}%
  \BibitemOpen
  \bibfield  {author} {\bibinfo {author} {\bibfnamefont {N.~S.}\ \bibnamefont
  {{Sugiyama}}}, \bibinfo {author} {\bibfnamefont {T.}~\bibnamefont
  {{Okumura}}}, \ and\ \bibinfo {author} {\bibfnamefont {D.~N.}\ \bibnamefont
  {{Spergel}}},\ }\href {\doibase 10.1093/mnras/stx3362} {\bibfield  {journal}
  {\bibinfo  {journal} {\mnras}\ }\textbf {\bibinfo {volume} {475}},\ \bibinfo
  {pages} {3764} (\bibinfo {year} {2018})},\ \Eprint
  {http://arxiv.org/abs/1705.07449} {arXiv:1705.07449 [astro-ph.CO]}
  \BibitemShut {NoStop}%
\bibitem [{\citenamefont {{Planck Collaboration}}\ \emph
  {et~al.}(2018)\citenamefont {{Planck Collaboration}}, \citenamefont
  {{Aghanim}}, \citenamefont {{Akrami}}, \citenamefont {{Ashdown}},
  \citenamefont {{Aumont}}, \citenamefont {{Baccigalupi}}, \citenamefont
  {{Ballardini}}, \citenamefont {{Banday}}, \citenamefont {{Barreiro}},
  \citenamefont {{Bartolo}}, \citenamefont {{Basak}}, \citenamefont {{Battye}},
  \citenamefont {{Benabed}}, \citenamefont {{Bernard}}, \citenamefont
  {{Bersanelli}}, \citenamefont {{Bielewicz}}, \citenamefont {{Bond}},
  \citenamefont {{Borrill}}, \citenamefont {{Bouchet}}, \citenamefont
  {{Burigana}}, \citenamefont {{Calabrese}}, \citenamefont {{Carron}},
  \citenamefont {{Chiang}}, \citenamefont {{Comis}}, \citenamefont
  {{Contreras}}, \citenamefont {{Crill}}, \citenamefont {{Curto}},
  \citenamefont {{Cuttaia}}, \citenamefont {{de Bernardis}}, \citenamefont {{de
  Rosa}}, \citenamefont {{de Zotti}}, \citenamefont {{Delabrouille}},
  \citenamefont {{Di Valentino}}, \citenamefont {{Dickinson}}, \citenamefont
  {{Diego}}, \citenamefont {{Dor{\'e}}}, \citenamefont {{Ducout}},
  \citenamefont {{Dupac}}, \citenamefont {{Elsner}}, \citenamefont
  {{En{\ss}lin}}, \citenamefont {{Eriksen}}, \citenamefont {{Falgarone}},
  \citenamefont {{Fantaye}}, \citenamefont {{Finelli}}, \citenamefont
  {{Forastieri}}, \citenamefont {{Frailis}}, \citenamefont {{Fraisse}},
  \citenamefont {{Franceschi}}, \citenamefont {{Frolov}}, \citenamefont
  {{Galeotta}}, \citenamefont {{Galli}}, \citenamefont {{Ganga}}, \citenamefont
  {{Gerbino}}, \citenamefont {{G{\'o}rski}}, \citenamefont {{Gruppuso}},
  \citenamefont {{Gudmundsson}}, \citenamefont {{Hand ley}}, \citenamefont
  {{Hansen}}, \citenamefont {{Herranz}}, \citenamefont {{Hivon}}, \citenamefont
  {{Huang}}, \citenamefont {{Jaffe}}, \citenamefont {{Keih{\"a}nen}},
  \citenamefont {{Keskitalo}}, \citenamefont {{Kiiveri}}, \citenamefont
  {{Kim}}, \citenamefont {{Kisner}}, \citenamefont {{Krachmalnicoff}},
  \citenamefont {{Kunz}}, \citenamefont {{Kurki-Suonio}}, \citenamefont
  {{Lamarre}}, \citenamefont {{Lasenby}}, \citenamefont {{Lattanzi}},
  \citenamefont {{Lawrence}}, \citenamefont {{Le Jeune}}, \citenamefont
  {{Levrier}}, \citenamefont {{Liguori}}, \citenamefont {{Lilje}},
  \citenamefont {{Lindholm}}, \citenamefont {{L{\'o}pez-Caniego}},
  \citenamefont {{Lubin}}, \citenamefont {{Ma}}, \citenamefont
  {{Mac{\'\i}as-P{\'e}rez}}, \citenamefont {{Maggio}}, \citenamefont {{Maino}},
  \citenamefont {{Mand olesi}}, \citenamefont {{Mangilli}}, \citenamefont
  {{Martin}}, \citenamefont {{Mart{\'\i}nez-Gonz{\'a}lez}}, \citenamefont
  {{Matarrese}}, \citenamefont {{Mauri}}, \citenamefont {{McEwen}},
  \citenamefont {{Melchiorri}}, \citenamefont {{Mennella}}, \citenamefont
  {{Migliaccio}}, \citenamefont {{Miville-Desch{\^e}nes}}, \citenamefont
  {{Molinari}}, \citenamefont {{Moneti}}, \citenamefont {{Montier}},
  \citenamefont {{Morgante}}, \citenamefont {{Natoli}}, \citenamefont
  {{Oxborrow}}, \citenamefont {{Pagano}}, \citenamefont {{Paoletti}},
  \citenamefont {{Partridge}}, \citenamefont {{Perdereau}}, \citenamefont
  {{Perotto}}, \citenamefont {{Pettorino}}, \citenamefont {{Piacentini}},
  \citenamefont {{Plaszczynski}}, \citenamefont {{Polastri}}, \citenamefont
  {{Polenta}}, \citenamefont {{Rachen}}, \citenamefont {{Racine}},
  \citenamefont {{Reinecke}}, \citenamefont {{Remazeilles}}, \citenamefont
  {{Renzi}}, \citenamefont {{Rocha}}, \citenamefont {{Roudier}}, \citenamefont
  {{Ruiz-Granados}}, \citenamefont {{Sandri}}, \citenamefont {{Savelainen}},
  \citenamefont {{Scott}}, \citenamefont {{Sirignano}}, \citenamefont
  {{Sirri}}, \citenamefont {{Spencer}}, \citenamefont {{Stanco}}, \citenamefont
  {{Sunyaev}}, \citenamefont {{Tauber}}, \citenamefont {{Tavagnacco}},
  \citenamefont {{Tenti}}, \citenamefont {{Toffolatti}}, \citenamefont
  {{Tomasi}}, \citenamefont {{Tristram}}, \citenamefont {{Trombetti}},
  \citenamefont {{Valiviita}}, \citenamefont {{Van Tent}}, \citenamefont
  {{Vielva}}, \citenamefont {{Villa}}, \citenamefont {{Vittorio}},
  \citenamefont {{Wandelt}}, \citenamefont {{Wehus}}, \citenamefont
  {{Zacchei}},\ and\ \citenamefont {{Zonca}}}]{PlanckkSZ18}%
  \BibitemOpen
  \bibfield  {author} {\bibinfo {author} {\bibnamefont {{Planck
  Collaboration}}}, \bibinfo {author} {\bibfnamefont {N.}~\bibnamefont
  {{Aghanim}}}, \bibinfo {author} {\bibfnamefont {Y.}~\bibnamefont {{Akrami}}},
  \bibinfo {author} {\bibfnamefont {M.}~\bibnamefont {{Ashdown}}}, \bibinfo
  {author} {\bibfnamefont {J.}~\bibnamefont {{Aumont}}}, \bibinfo {author}
  {\bibfnamefont {C.}~\bibnamefont {{Baccigalupi}}}, \bibinfo {author}
  {\bibfnamefont {M.}~\bibnamefont {{Ballardini}}}, \bibinfo {author}
  {\bibfnamefont {A.~J.}\ \bibnamefont {{Banday}}}, \bibinfo {author}
  {\bibfnamefont {R.~B.}\ \bibnamefont {{Barreiro}}}, \bibinfo {author}
  {\bibfnamefont {N.}~\bibnamefont {{Bartolo}}}, \bibinfo {author}
  {\bibfnamefont {S.}~\bibnamefont {{Basak}}}, \bibinfo {author} {\bibfnamefont
  {R.}~\bibnamefont {{Battye}}}, \bibinfo {author} {\bibfnamefont
  {K.}~\bibnamefont {{Benabed}}}, \bibinfo {author} {\bibfnamefont {J.~P.}\
  \bibnamefont {{Bernard}}}, \bibinfo {author} {\bibfnamefont {M.}~\bibnamefont
  {{Bersanelli}}}, \bibinfo {author} {\bibfnamefont {P.}~\bibnamefont
  {{Bielewicz}}}, \bibinfo {author} {\bibfnamefont {J.~R.}\ \bibnamefont
  {{Bond}}}, \bibinfo {author} {\bibfnamefont {J.}~\bibnamefont {{Borrill}}},
  \bibinfo {author} {\bibfnamefont {F.~R.}\ \bibnamefont {{Bouchet}}}, \bibinfo
  {author} {\bibfnamefont {C.}~\bibnamefont {{Burigana}}}, \bibinfo {author}
  {\bibfnamefont {E.}~\bibnamefont {{Calabrese}}}, \bibinfo {author}
  {\bibfnamefont {J.}~\bibnamefont {{Carron}}}, \bibinfo {author}
  {\bibfnamefont {H.~C.}\ \bibnamefont {{Chiang}}}, \bibinfo {author}
  {\bibfnamefont {B.}~\bibnamefont {{Comis}}}, \bibinfo {author} {\bibfnamefont
  {D.}~\bibnamefont {{Contreras}}}, \bibinfo {author} {\bibfnamefont {B.~P.}\
  \bibnamefont {{Crill}}}, \bibinfo {author} {\bibfnamefont {A.}~\bibnamefont
  {{Curto}}}, \bibinfo {author} {\bibfnamefont {F.}~\bibnamefont {{Cuttaia}}},
  \bibinfo {author} {\bibfnamefont {P.}~\bibnamefont {{de Bernardis}}},
  \bibinfo {author} {\bibfnamefont {A.}~\bibnamefont {{de Rosa}}}, \bibinfo
  {author} {\bibfnamefont {G.}~\bibnamefont {{de Zotti}}}, \bibinfo {author}
  {\bibfnamefont {J.}~\bibnamefont {{Delabrouille}}}, \bibinfo {author}
  {\bibfnamefont {E.}~\bibnamefont {{Di Valentino}}}, \bibinfo {author}
  {\bibfnamefont {C.}~\bibnamefont {{Dickinson}}}, \bibinfo {author}
  {\bibfnamefont {J.~M.}\ \bibnamefont {{Diego}}}, \bibinfo {author}
  {\bibfnamefont {O.}~\bibnamefont {{Dor{\'e}}}}, \bibinfo {author}
  {\bibfnamefont {A.}~\bibnamefont {{Ducout}}}, \bibinfo {author}
  {\bibfnamefont {X.}~\bibnamefont {{Dupac}}}, \bibinfo {author} {\bibfnamefont
  {F.}~\bibnamefont {{Elsner}}}, \bibinfo {author} {\bibfnamefont {T.~A.}\
  \bibnamefont {{En{\ss}lin}}}, \bibinfo {author} {\bibfnamefont {H.~K.}\
  \bibnamefont {{Eriksen}}}, \bibinfo {author} {\bibfnamefont {E.}~\bibnamefont
  {{Falgarone}}}, \bibinfo {author} {\bibfnamefont {Y.}~\bibnamefont
  {{Fantaye}}}, \bibinfo {author} {\bibfnamefont {F.}~\bibnamefont
  {{Finelli}}}, \bibinfo {author} {\bibfnamefont {F.}~\bibnamefont
  {{Forastieri}}}, \bibinfo {author} {\bibfnamefont {M.}~\bibnamefont
  {{Frailis}}}, \bibinfo {author} {\bibfnamefont {A.~A.}\ \bibnamefont
  {{Fraisse}}}, \bibinfo {author} {\bibfnamefont {E.}~\bibnamefont
  {{Franceschi}}}, \bibinfo {author} {\bibfnamefont {A.}~\bibnamefont
  {{Frolov}}}, \bibinfo {author} {\bibfnamefont {S.}~\bibnamefont
  {{Galeotta}}}, \bibinfo {author} {\bibfnamefont {S.}~\bibnamefont {{Galli}}},
  \bibinfo {author} {\bibfnamefont {K.}~\bibnamefont {{Ganga}}}, \bibinfo
  {author} {\bibfnamefont {M.}~\bibnamefont {{Gerbino}}}, \bibinfo {author}
  {\bibfnamefont {K.~M.}\ \bibnamefont {{G{\'o}rski}}}, \bibinfo {author}
  {\bibfnamefont {A.}~\bibnamefont {{Gruppuso}}}, \bibinfo {author}
  {\bibfnamefont {J.~E.}\ \bibnamefont {{Gudmundsson}}}, \bibinfo {author}
  {\bibfnamefont {W.}~\bibnamefont {{Hand ley}}}, \bibinfo {author}
  {\bibfnamefont {F.~K.}\ \bibnamefont {{Hansen}}}, \bibinfo {author}
  {\bibfnamefont {D.}~\bibnamefont {{Herranz}}}, \bibinfo {author}
  {\bibfnamefont {E.}~\bibnamefont {{Hivon}}}, \bibinfo {author} {\bibfnamefont
  {Z.}~\bibnamefont {{Huang}}}, \bibinfo {author} {\bibfnamefont {A.~H.}\
  \bibnamefont {{Jaffe}}}, \bibinfo {author} {\bibfnamefont {E.}~\bibnamefont
  {{Keih{\"a}nen}}}, \bibinfo {author} {\bibfnamefont {R.}~\bibnamefont
  {{Keskitalo}}}, \bibinfo {author} {\bibfnamefont {K.}~\bibnamefont
  {{Kiiveri}}}, \bibinfo {author} {\bibfnamefont {J.}~\bibnamefont {{Kim}}},
  \bibinfo {author} {\bibfnamefont {T.~S.}\ \bibnamefont {{Kisner}}}, \bibinfo
  {author} {\bibfnamefont {N.}~\bibnamefont {{Krachmalnicoff}}}, \bibinfo
  {author} {\bibfnamefont {M.}~\bibnamefont {{Kunz}}}, \bibinfo {author}
  {\bibfnamefont {H.}~\bibnamefont {{Kurki-Suonio}}}, \bibinfo {author}
  {\bibfnamefont {J.~M.}\ \bibnamefont {{Lamarre}}}, \bibinfo {author}
  {\bibfnamefont {A.}~\bibnamefont {{Lasenby}}}, \bibinfo {author}
  {\bibfnamefont {M.}~\bibnamefont {{Lattanzi}}}, \bibinfo {author}
  {\bibfnamefont {C.~R.}\ \bibnamefont {{Lawrence}}}, \bibinfo {author}
  {\bibfnamefont {M.}~\bibnamefont {{Le Jeune}}}, \bibinfo {author}
  {\bibfnamefont {F.}~\bibnamefont {{Levrier}}}, \bibinfo {author}
  {\bibfnamefont {M.}~\bibnamefont {{Liguori}}}, \bibinfo {author}
  {\bibfnamefont {P.~B.}\ \bibnamefont {{Lilje}}}, \bibinfo {author}
  {\bibfnamefont {V.}~\bibnamefont {{Lindholm}}}, \bibinfo {author}
  {\bibfnamefont {M.}~\bibnamefont {{L{\'o}pez-Caniego}}}, \bibinfo {author}
  {\bibfnamefont {P.~M.}\ \bibnamefont {{Lubin}}}, \bibinfo {author}
  {\bibfnamefont {Y.~Z.}\ \bibnamefont {{Ma}}}, \bibinfo {author}
  {\bibfnamefont {J.~F.}\ \bibnamefont {{Mac{\'\i}as-P{\'e}rez}}}, \bibinfo
  {author} {\bibfnamefont {G.}~\bibnamefont {{Maggio}}}, \bibinfo {author}
  {\bibfnamefont {D.}~\bibnamefont {{Maino}}}, \bibinfo {author} {\bibfnamefont
  {N.}~\bibnamefont {{Mand olesi}}}, \bibinfo {author} {\bibfnamefont
  {A.}~\bibnamefont {{Mangilli}}}, \bibinfo {author} {\bibfnamefont {P.~G.}\
  \bibnamefont {{Martin}}}, \bibinfo {author} {\bibfnamefont {E.}~\bibnamefont
  {{Mart{\'\i}nez-Gonz{\'a}lez}}}, \bibinfo {author} {\bibfnamefont
  {S.}~\bibnamefont {{Matarrese}}}, \bibinfo {author} {\bibfnamefont
  {N.}~\bibnamefont {{Mauri}}}, \bibinfo {author} {\bibfnamefont {J.~D.}\
  \bibnamefont {{McEwen}}}, \bibinfo {author} {\bibfnamefont {A.}~\bibnamefont
  {{Melchiorri}}}, \bibinfo {author} {\bibfnamefont {A.}~\bibnamefont
  {{Mennella}}}, \bibinfo {author} {\bibfnamefont {M.}~\bibnamefont
  {{Migliaccio}}}, \bibinfo {author} {\bibfnamefont {M.~A.}\ \bibnamefont
  {{Miville-Desch{\^e}nes}}}, \bibinfo {author} {\bibfnamefont
  {D.}~\bibnamefont {{Molinari}}}, \bibinfo {author} {\bibfnamefont
  {A.}~\bibnamefont {{Moneti}}}, \bibinfo {author} {\bibfnamefont
  {L.}~\bibnamefont {{Montier}}}, \bibinfo {author} {\bibfnamefont
  {G.}~\bibnamefont {{Morgante}}}, \bibinfo {author} {\bibfnamefont
  {P.}~\bibnamefont {{Natoli}}}, \bibinfo {author} {\bibfnamefont {C.~A.}\
  \bibnamefont {{Oxborrow}}}, \bibinfo {author} {\bibfnamefont
  {L.}~\bibnamefont {{Pagano}}}, \bibinfo {author} {\bibfnamefont
  {D.}~\bibnamefont {{Paoletti}}}, \bibinfo {author} {\bibfnamefont
  {B.}~\bibnamefont {{Partridge}}}, \bibinfo {author} {\bibfnamefont
  {O.}~\bibnamefont {{Perdereau}}}, \bibinfo {author} {\bibfnamefont
  {L.}~\bibnamefont {{Perotto}}}, \bibinfo {author} {\bibfnamefont
  {V.}~\bibnamefont {{Pettorino}}}, \bibinfo {author} {\bibfnamefont
  {F.}~\bibnamefont {{Piacentini}}}, \bibinfo {author} {\bibfnamefont
  {S.}~\bibnamefont {{Plaszczynski}}}, \bibinfo {author} {\bibfnamefont
  {L.}~\bibnamefont {{Polastri}}}, \bibinfo {author} {\bibfnamefont
  {G.}~\bibnamefont {{Polenta}}}, \bibinfo {author} {\bibfnamefont {J.~P.}\
  \bibnamefont {{Rachen}}}, \bibinfo {author} {\bibfnamefont {B.}~\bibnamefont
  {{Racine}}}, \bibinfo {author} {\bibfnamefont {M.}~\bibnamefont
  {{Reinecke}}}, \bibinfo {author} {\bibfnamefont {M.}~\bibnamefont
  {{Remazeilles}}}, \bibinfo {author} {\bibfnamefont {A.}~\bibnamefont
  {{Renzi}}}, \bibinfo {author} {\bibfnamefont {G.}~\bibnamefont {{Rocha}}},
  \bibinfo {author} {\bibfnamefont {G.}~\bibnamefont {{Roudier}}}, \bibinfo
  {author} {\bibfnamefont {B.}~\bibnamefont {{Ruiz-Granados}}}, \bibinfo
  {author} {\bibfnamefont {M.}~\bibnamefont {{Sandri}}}, \bibinfo {author}
  {\bibfnamefont {M.}~\bibnamefont {{Savelainen}}}, \bibinfo {author}
  {\bibfnamefont {D.}~\bibnamefont {{Scott}}}, \bibinfo {author} {\bibfnamefont
  {C.}~\bibnamefont {{Sirignano}}}, \bibinfo {author} {\bibfnamefont
  {G.}~\bibnamefont {{Sirri}}}, \bibinfo {author} {\bibfnamefont {L.~D.}\
  \bibnamefont {{Spencer}}}, \bibinfo {author} {\bibfnamefont {L.}~\bibnamefont
  {{Stanco}}}, \bibinfo {author} {\bibfnamefont {R.}~\bibnamefont {{Sunyaev}}},
  \bibinfo {author} {\bibfnamefont {J.~A.}\ \bibnamefont {{Tauber}}}, \bibinfo
  {author} {\bibfnamefont {D.}~\bibnamefont {{Tavagnacco}}}, \bibinfo {author}
  {\bibfnamefont {M.}~\bibnamefont {{Tenti}}}, \bibinfo {author} {\bibfnamefont
  {L.}~\bibnamefont {{Toffolatti}}}, \bibinfo {author} {\bibfnamefont
  {M.}~\bibnamefont {{Tomasi}}}, \bibinfo {author} {\bibfnamefont
  {M.}~\bibnamefont {{Tristram}}}, \bibinfo {author} {\bibfnamefont
  {T.}~\bibnamefont {{Trombetti}}}, \bibinfo {author} {\bibfnamefont
  {J.}~\bibnamefont {{Valiviita}}}, \bibinfo {author} {\bibfnamefont
  {F.}~\bibnamefont {{Van Tent}}}, \bibinfo {author} {\bibfnamefont
  {P.}~\bibnamefont {{Vielva}}}, \bibinfo {author} {\bibfnamefont
  {F.}~\bibnamefont {{Villa}}}, \bibinfo {author} {\bibfnamefont
  {N.}~\bibnamefont {{Vittorio}}}, \bibinfo {author} {\bibfnamefont {B.~D.}\
  \bibnamefont {{Wandelt}}}, \bibinfo {author} {\bibfnamefont {I.~K.}\
  \bibnamefont {{Wehus}}}, \bibinfo {author} {\bibfnamefont {A.}~\bibnamefont
  {{Zacchei}}}, \ and\ \bibinfo {author} {\bibfnamefont {A.}~\bibnamefont
  {{Zonca}}},\ }\href {\doibase 10.1051/0004-6361/201731489} {\bibfield
  {journal} {\bibinfo  {journal} {\aap}\ }\textbf {\bibinfo {volume} {617}},\
  \bibinfo {eid} {A48} (\bibinfo {year} {2018})},\ \Eprint
  {http://arxiv.org/abs/1707.00132} {arXiv:1707.00132 [astro-ph.CO]}
  \BibitemShut {NoStop}%
\bibitem [{\citenamefont {{Flender}}\ \emph {et~al.}(2016)\citenamefont
  {{Flender}}, \citenamefont {{Bleem}}, \citenamefont {{Finkel}}, \citenamefont
  {{Habib}}, \citenamefont {{Heitmann}},\ and\ \citenamefont
  {{Holder}}}]{Flender16}%
  \BibitemOpen
  \bibfield  {author} {\bibinfo {author} {\bibfnamefont {S.}~\bibnamefont
  {{Flender}}}, \bibinfo {author} {\bibfnamefont {L.}~\bibnamefont {{Bleem}}},
  \bibinfo {author} {\bibfnamefont {H.}~\bibnamefont {{Finkel}}}, \bibinfo
  {author} {\bibfnamefont {S.}~\bibnamefont {{Habib}}}, \bibinfo {author}
  {\bibfnamefont {K.}~\bibnamefont {{Heitmann}}}, \ and\ \bibinfo {author}
  {\bibfnamefont {G.}~\bibnamefont {{Holder}}},\ }\href {\doibase
  10.3847/0004-637X/823/2/98} {\bibfield  {journal} {\bibinfo  {journal}
  {\apj}\ }\textbf {\bibinfo {volume} {823}},\ \bibinfo {eid} {98} (\bibinfo
  {year} {2016})},\ \Eprint {http://arxiv.org/abs/1511.02843} {arXiv:1511.02843
  [astro-ph.CO]} \BibitemShut {NoStop}%
\bibitem [{\citenamefont {{Sugiyama}}\ \emph {et~al.}(2017)\citenamefont
  {{Sugiyama}}, \citenamefont {{Okumura}},\ and\ \citenamefont
  {{Spergel}}}]{Sugiyama2017}%
  \BibitemOpen
  \bibfield  {author} {\bibinfo {author} {\bibfnamefont {N.~S.}\ \bibnamefont
  {{Sugiyama}}}, \bibinfo {author} {\bibfnamefont {T.}~\bibnamefont
  {{Okumura}}}, \ and\ \bibinfo {author} {\bibfnamefont {D.~N.}\ \bibnamefont
  {{Spergel}}},\ }\href {\doibase 10.1088/1475-7516/2017/01/057} {\bibfield
  {journal} {\bibinfo  {journal} {\jcap}\ }\textbf {\bibinfo {volume} {2017}},\
  \bibinfo {eid} {057} (\bibinfo {year} {2017})},\ \Eprint
  {http://arxiv.org/abs/1606.06367} {arXiv:1606.06367 [astro-ph.CO]}
  \BibitemShut {NoStop}%
\bibitem [{\citenamefont {{Ferreira}}\ \emph {et~al.}(1999)\citenamefont
  {{Ferreira}}, \citenamefont {{Juszkiewicz}}, \citenamefont {{Feldman}},
  \citenamefont {{Davis}},\ and\ \citenamefont {{Jaffe}}}]{Ferreira1999}%
  \BibitemOpen
  \bibfield  {author} {\bibinfo {author} {\bibfnamefont {P.~G.}\ \bibnamefont
  {{Ferreira}}}, \bibinfo {author} {\bibfnamefont {R.}~\bibnamefont
  {{Juszkiewicz}}}, \bibinfo {author} {\bibfnamefont {H.~A.}\ \bibnamefont
  {{Feldman}}}, \bibinfo {author} {\bibfnamefont {M.}~\bibnamefont {{Davis}}},
  \ and\ \bibinfo {author} {\bibfnamefont {A.~H.}\ \bibnamefont {{Jaffe}}},\
  }\href {\doibase 10.1086/311959} {\bibfield  {journal} {\bibinfo  {journal}
  {\apjl}\ }\textbf {\bibinfo {volume} {515}},\ \bibinfo {pages} {L1} (\bibinfo
  {year} {1999})},\ \Eprint {http://arxiv.org/abs/astro-ph/9812456}
  {arXiv:astro-ph/9812456 [astro-ph]} \BibitemShut {NoStop}%
\bibitem [{\citenamefont {{Sugiyama}}\ \emph {et~al.}(2016)\citenamefont
  {{Sugiyama}}, \citenamefont {{Okumura}},\ and\ \citenamefont
  {{Spergel}}}]{Sugiyama2016}%
  \BibitemOpen
  \bibfield  {author} {\bibinfo {author} {\bibfnamefont {N.~S.}\ \bibnamefont
  {{Sugiyama}}}, \bibinfo {author} {\bibfnamefont {T.}~\bibnamefont
  {{Okumura}}}, \ and\ \bibinfo {author} {\bibfnamefont {D.~N.}\ \bibnamefont
  {{Spergel}}},\ }\href {\doibase 10.1088/1475-7516/2016/07/001} {\bibfield
  {journal} {\bibinfo  {journal} {\jcap}\ }\textbf {\bibinfo {volume} {2016}},\
  \bibinfo {eid} {001} (\bibinfo {year} {2016})},\ \Eprint
  {http://arxiv.org/abs/1509.08232} {arXiv:1509.08232 [astro-ph.CO]}
  \BibitemShut {NoStop}%
\bibitem [{\citenamefont {{Okumura}}\ \emph {et~al.}(2014)\citenamefont
  {{Okumura}}, \citenamefont {{Seljak}}, \citenamefont {{Vlah}},\ and\
  \citenamefont {{Desjacques}}}]{Okumura2014}%
  \BibitemOpen
  \bibfield  {author} {\bibinfo {author} {\bibfnamefont {T.}~\bibnamefont
  {{Okumura}}}, \bibinfo {author} {\bibfnamefont {U.}~\bibnamefont {{Seljak}}},
  \bibinfo {author} {\bibfnamefont {Z.}~\bibnamefont {{Vlah}}}, \ and\ \bibinfo
  {author} {\bibfnamefont {V.}~\bibnamefont {{Desjacques}}},\ }\href {\doibase
  10.1088/1475-7516/2014/05/003} {\bibfield  {journal} {\bibinfo  {journal}
  {\jcap}\ }\textbf {\bibinfo {volume} {2014}},\ \bibinfo {eid} {003} (\bibinfo
  {year} {2014})},\ \Eprint {http://arxiv.org/abs/1312.4214} {arXiv:1312.4214
  [astro-ph.CO]} \BibitemShut {NoStop}%
\bibitem [{\citenamefont {{De Felice}}\ and\ \citenamefont
  {{Tsujikawa}}(2010)}]{fRreview}%
  \BibitemOpen
  \bibfield  {author} {\bibinfo {author} {\bibfnamefont {A.}~\bibnamefont {{De
  Felice}}}\ and\ \bibinfo {author} {\bibfnamefont {S.}~\bibnamefont
  {{Tsujikawa}}},\ }\href {\doibase 10.12942/lrr-2010-3} {\bibfield  {journal}
  {\bibinfo  {journal} {Living Reviews in Relativity}\ }\textbf {\bibinfo
  {volume} {13}},\ \bibinfo {eid} {3} (\bibinfo {year} {2010})},\ \Eprint
  {http://arxiv.org/abs/1002.4928} {arXiv:1002.4928 [gr-qc]} \BibitemShut
  {NoStop}%
\bibitem [{\citenamefont {{Dvali}}\ \emph {et~al.}(2000)\citenamefont
  {{Dvali}}, \citenamefont {{Gabadadze}},\ and\ \citenamefont
  {{Porrati}}}]{DGP}%
  \BibitemOpen
  \bibfield  {author} {\bibinfo {author} {\bibfnamefont {G.}~\bibnamefont
  {{Dvali}}}, \bibinfo {author} {\bibfnamefont {G.}~\bibnamefont
  {{Gabadadze}}}, \ and\ \bibinfo {author} {\bibfnamefont {M.}~\bibnamefont
  {{Porrati}}},\ }\href {\doibase 10.1016/S0370-2693(00)00669-9} {\bibfield
  {journal} {\bibinfo  {journal} {Physics Letters B}\ }\textbf {\bibinfo
  {volume} {485}},\ \bibinfo {pages} {208} (\bibinfo {year} {2000})},\ \Eprint
  {http://arxiv.org/abs/hep-th/0005016} {arXiv:hep-th/0005016 [hep-th]}
  \BibitemShut {NoStop}%
\bibitem [{\citenamefont {{Li}}\ \emph {et~al.}(2012)\citenamefont {{Li}},
  \citenamefont {{Zhao}}, \citenamefont {{Teyssier}},\ and\ \citenamefont
  {{Koyama}}}]{ECOSMOG}%
  \BibitemOpen
  \bibfield  {author} {\bibinfo {author} {\bibfnamefont {B.}~\bibnamefont
  {{Li}}}, \bibinfo {author} {\bibfnamefont {G.-B.}\ \bibnamefont {{Zhao}}},
  \bibinfo {author} {\bibfnamefont {R.}~\bibnamefont {{Teyssier}}}, \ and\
  \bibinfo {author} {\bibfnamefont {K.}~\bibnamefont {{Koyama}}},\ }\href
  {\doibase 10.1088/1475-7516/2012/01/051} {\bibfield  {journal} {\bibinfo
  {journal} {\jcap}\ }\textbf {\bibinfo {volume} {2012}},\ \bibinfo {eid} {051}
  (\bibinfo {year} {2012})},\ \Eprint {http://arxiv.org/abs/1110.1379}
  {arXiv:1110.1379 [astro-ph.CO]} \BibitemShut {NoStop}%
\bibitem [{\citenamefont {{Bose}}\ \emph {et~al.}(2017)\citenamefont {{Bose}},
  \citenamefont {{Li}}, \citenamefont {{Barreira}}, \citenamefont {{He}},
  \citenamefont {{Hellwing}}, \citenamefont {{Koyama}}, \citenamefont
  {{Llinares}},\ and\ \citenamefont {{Zhao}}}]{Bose2017}%
  \BibitemOpen
  \bibfield  {author} {\bibinfo {author} {\bibfnamefont {S.}~\bibnamefont
  {{Bose}}}, \bibinfo {author} {\bibfnamefont {B.}~\bibnamefont {{Li}}},
  \bibinfo {author} {\bibfnamefont {A.}~\bibnamefont {{Barreira}}}, \bibinfo
  {author} {\bibfnamefont {J.-h.}\ \bibnamefont {{He}}}, \bibinfo {author}
  {\bibfnamefont {W.~A.}\ \bibnamefont {{Hellwing}}}, \bibinfo {author}
  {\bibfnamefont {K.}~\bibnamefont {{Koyama}}}, \bibinfo {author}
  {\bibfnamefont {C.}~\bibnamefont {{Llinares}}}, \ and\ \bibinfo {author}
  {\bibfnamefont {G.-B.}\ \bibnamefont {{Zhao}}},\ }\href {\doibase
  10.1088/1475-7516/2017/02/050} {\bibfield  {journal} {\bibinfo  {journal}
  {\jcap}\ }\textbf {\bibinfo {volume} {2017}},\ \bibinfo {eid} {050} (\bibinfo
  {year} {2017})},\ \Eprint {http://arxiv.org/abs/1611.09375} {arXiv:1611.09375
  [astro-ph.CO]} \BibitemShut {NoStop}%
\bibitem [{\citenamefont {{Li}}\ \emph {et~al.}(2013)\citenamefont {{Li}},
  \citenamefont {{Zhao}},\ and\ \citenamefont {{Koyama}}}]{Li2013}%
  \BibitemOpen
  \bibfield  {author} {\bibinfo {author} {\bibfnamefont {B.}~\bibnamefont
  {{Li}}}, \bibinfo {author} {\bibfnamefont {G.-B.}\ \bibnamefont {{Zhao}}}, \
  and\ \bibinfo {author} {\bibfnamefont {K.}~\bibnamefont {{Koyama}}},\ }\href
  {\doibase 10.1088/1475-7516/2013/05/023} {\bibfield  {journal} {\bibinfo
  {journal} {\jcap}\ }\textbf {\bibinfo {volume} {2013}},\ \bibinfo {eid} {023}
  (\bibinfo {year} {2013})},\ \Eprint {http://arxiv.org/abs/1303.0008}
  {arXiv:1303.0008 [astro-ph.CO]} \BibitemShut {NoStop}%
\bibitem [{\citenamefont {{Barreira}}\ \emph {et~al.}(2015)\citenamefont
  {{Barreira}}, \citenamefont {{Bose}},\ and\ \citenamefont
  {{Li}}}]{Barreira2015}%
  \BibitemOpen
  \bibfield  {author} {\bibinfo {author} {\bibfnamefont {A.}~\bibnamefont
  {{Barreira}}}, \bibinfo {author} {\bibfnamefont {S.}~\bibnamefont {{Bose}}},
  \ and\ \bibinfo {author} {\bibfnamefont {B.}~\bibnamefont {{Li}}},\ }\href
  {\doibase 10.1088/1475-7516/2015/12/059} {\bibfield  {journal} {\bibinfo
  {journal} {\jcap}\ }\textbf {\bibinfo {volume} {2015}},\ \bibinfo {eid} {059}
  (\bibinfo {year} {2015})},\ \Eprint {http://arxiv.org/abs/1511.08200}
  {arXiv:1511.08200 [astro-ph.CO]} \BibitemShut {NoStop}%
\bibitem [{\citenamefont {Hinshaw}\ \emph {et~al.}(2013)\citenamefont
  {Hinshaw}, \citenamefont {Larson}, \citenamefont {Komatsu}, \citenamefont
  {Spergel}, \citenamefont {Bennett}, \citenamefont {Dunkley}, \citenamefont
  {Nolta}, \citenamefont {Halpern}, \citenamefont {Hill}, \citenamefont
  {Odegard}, \citenamefont {Page}, \citenamefont {Smith}, \citenamefont
  {Weiland}, \citenamefont {Gold}, \citenamefont {Jarosik}, \citenamefont
  {Kogut}, \citenamefont {Limon}, \citenamefont {Meyer}, \citenamefont
  {Tucker}, \citenamefont {Wollack},\ and\ \citenamefont {Wright}}]{WMAP9}%
  \BibitemOpen
  \bibfield  {author} {\bibinfo {author} {\bibfnamefont {G.}~\bibnamefont
  {Hinshaw}}, \bibinfo {author} {\bibfnamefont {D.}~\bibnamefont {Larson}},
  \bibinfo {author} {\bibfnamefont {E.}~\bibnamefont {Komatsu}}, \bibinfo
  {author} {\bibfnamefont {D.~N.}\ \bibnamefont {Spergel}}, \bibinfo {author}
  {\bibfnamefont {C.~L.}\ \bibnamefont {Bennett}}, \bibinfo {author}
  {\bibfnamefont {J.}~\bibnamefont {Dunkley}}, \bibinfo {author} {\bibfnamefont
  {M.~R.}\ \bibnamefont {Nolta}}, \bibinfo {author} {\bibfnamefont
  {M.}~\bibnamefont {Halpern}}, \bibinfo {author} {\bibfnamefont {R.~S.}\
  \bibnamefont {Hill}}, \bibinfo {author} {\bibfnamefont {N.}~\bibnamefont
  {Odegard}}, \bibinfo {author} {\bibfnamefont {L.}~\bibnamefont {Page}},
  \bibinfo {author} {\bibfnamefont {K.~M.}\ \bibnamefont {Smith}}, \bibinfo
  {author} {\bibfnamefont {J.~L.}\ \bibnamefont {Weiland}}, \bibinfo {author}
  {\bibfnamefont {B.}~\bibnamefont {Gold}}, \bibinfo {author} {\bibfnamefont
  {N.}~\bibnamefont {Jarosik}}, \bibinfo {author} {\bibfnamefont
  {A.}~\bibnamefont {Kogut}}, \bibinfo {author} {\bibfnamefont
  {M.}~\bibnamefont {Limon}}, \bibinfo {author} {\bibfnamefont {S.~S.}\
  \bibnamefont {Meyer}}, \bibinfo {author} {\bibfnamefont {G.~S.}\ \bibnamefont
  {Tucker}}, \bibinfo {author} {\bibfnamefont {E.}~\bibnamefont {Wollack}}, \
  and\ \bibinfo {author} {\bibfnamefont {E.~L.}\ \bibnamefont {Wright}},\
  }\href {\doibase 10.1088/0067-0049/208/2/19} {\bibfield  {journal} {\bibinfo
  {journal} {The Astrophysical Journal Supplement Series}\ }\textbf {\bibinfo
  {volume} {208}},\ \bibinfo {pages} {19} (\bibinfo {year} {2013})}\BibitemShut
  {NoStop}%
\bibitem [{\citenamefont {{Manera}}\ \emph {et~al.}(2013)\citenamefont
  {{Manera}}, \citenamefont {{Scoccimarro}}, \citenamefont {{Percival}},
  \citenamefont {{Samushia}}, \citenamefont {{McBride}}, \citenamefont
  {{Ross}}, \citenamefont {{Sheth}}, \citenamefont {{White}}, \citenamefont
  {{Reid}}, \citenamefont {{S{\'a}nchez}}, \citenamefont {{de Putter}},
  \citenamefont {{Xu}}, \citenamefont {{Berlind}}, \citenamefont {{Brinkmann}},
  \citenamefont {{Maraston}}, \citenamefont {{Nichol}}, \citenamefont
  {{Montesano}}, \citenamefont {{Padmanabhan}}, \citenamefont {{Skibba}},
  \citenamefont {{Tojeiro}},\ and\ \citenamefont {{Weaver}}}]{Manera2013}%
  \BibitemOpen
  \bibfield  {author} {\bibinfo {author} {\bibfnamefont {M.}~\bibnamefont
  {{Manera}}}, \bibinfo {author} {\bibfnamefont {R.}~\bibnamefont
  {{Scoccimarro}}}, \bibinfo {author} {\bibfnamefont {W.~J.}\ \bibnamefont
  {{Percival}}}, \bibinfo {author} {\bibfnamefont {L.}~\bibnamefont
  {{Samushia}}}, \bibinfo {author} {\bibfnamefont {C.~K.}\ \bibnamefont
  {{McBride}}}, \bibinfo {author} {\bibfnamefont {A.~J.}\ \bibnamefont
  {{Ross}}}, \bibinfo {author} {\bibfnamefont {R.~K.}\ \bibnamefont {{Sheth}}},
  \bibinfo {author} {\bibfnamefont {M.}~\bibnamefont {{White}}}, \bibinfo
  {author} {\bibfnamefont {B.~A.}\ \bibnamefont {{Reid}}}, \bibinfo {author}
  {\bibfnamefont {A.~G.}\ \bibnamefont {{S{\'a}nchez}}}, \bibinfo {author}
  {\bibfnamefont {R.}~\bibnamefont {{de Putter}}}, \bibinfo {author}
  {\bibfnamefont {X.}~\bibnamefont {{Xu}}}, \bibinfo {author} {\bibfnamefont
  {A.~A.}\ \bibnamefont {{Berlind}}}, \bibinfo {author} {\bibfnamefont
  {J.}~\bibnamefont {{Brinkmann}}}, \bibinfo {author} {\bibfnamefont
  {C.}~\bibnamefont {{Maraston}}}, \bibinfo {author} {\bibfnamefont
  {B.}~\bibnamefont {{Nichol}}}, \bibinfo {author} {\bibfnamefont
  {F.}~\bibnamefont {{Montesano}}}, \bibinfo {author} {\bibfnamefont
  {N.}~\bibnamefont {{Padmanabhan}}}, \bibinfo {author} {\bibfnamefont {R.~A.}\
  \bibnamefont {{Skibba}}}, \bibinfo {author} {\bibfnamefont {R.}~\bibnamefont
  {{Tojeiro}}}, \ and\ \bibinfo {author} {\bibfnamefont {B.~A.}\ \bibnamefont
  {{Weaver}}},\ }\href {\doibase 10.1093/mnras/sts084} {\bibfield  {journal}
  {\bibinfo  {journal} {\mnras}\ }\textbf {\bibinfo {volume} {428}},\ \bibinfo
  {pages} {1036} (\bibinfo {year} {2013})},\ \Eprint
  {http://arxiv.org/abs/1203.6609} {arXiv:1203.6609 [astro-ph.CO]} \BibitemShut
  {NoStop}%
\bibitem [{\citenamefont {{DESI Collaboration}}\ \emph
  {et~al.}(2016)\citenamefont {{DESI Collaboration}}, \citenamefont
  {{Aghamousa}}, \citenamefont {{Aguilar}}, \citenamefont {{Ahlen}},
  \citenamefont {{Alam}}, \citenamefont {{Allen}}, \citenamefont {{Allende
  Prieto}}, \citenamefont {{Annis}}, \citenamefont {{Bailey}}, \citenamefont
  {{Balland}},\ and\ \citenamefont {et~al.}}]{DESI16I}%
  \BibitemOpen
  \bibfield  {author} {\bibinfo {author} {\bibnamefont {{DESI Collaboration}}},
  \bibinfo {author} {\bibfnamefont {A.}~\bibnamefont {{Aghamousa}}}, \bibinfo
  {author} {\bibfnamefont {J.}~\bibnamefont {{Aguilar}}}, \bibinfo {author}
  {\bibfnamefont {S.}~\bibnamefont {{Ahlen}}}, \bibinfo {author} {\bibfnamefont
  {S.}~\bibnamefont {{Alam}}}, \bibinfo {author} {\bibfnamefont {L.~E.}\
  \bibnamefont {{Allen}}}, \bibinfo {author} {\bibfnamefont {C.}~\bibnamefont
  {{Allende Prieto}}}, \bibinfo {author} {\bibfnamefont {J.}~\bibnamefont
  {{Annis}}}, \bibinfo {author} {\bibfnamefont {S.}~\bibnamefont {{Bailey}}},
  \bibinfo {author} {\bibfnamefont {C.}~\bibnamefont {{Balland}}}, \ and\
  \bibinfo {author} {\bibnamefont {et~al.}},\ }\href@noop {} {\bibfield
  {journal} {\bibinfo  {journal} {ArXiv e-prints}\ } (\bibinfo {year}
  {2016})},\ \Eprint {http://arxiv.org/abs/1611.00036} {arXiv:1611.00036
  [astro-ph.IM]} \BibitemShut {NoStop}%
\bibitem [{\citenamefont {{Behroozi}}\ \emph {et~al.}(2013)\citenamefont
  {{Behroozi}}, \citenamefont {{Wechsler}},\ and\ \citenamefont
  {{Wu}}}]{Rockstar}%
  \BibitemOpen
  \bibfield  {author} {\bibinfo {author} {\bibfnamefont {P.~S.}\ \bibnamefont
  {{Behroozi}}}, \bibinfo {author} {\bibfnamefont {R.~H.}\ \bibnamefont
  {{Wechsler}}}, \ and\ \bibinfo {author} {\bibfnamefont {H.-Y.}\ \bibnamefont
  {{Wu}}},\ }\href {\doibase 10.1088/0004-637X/762/2/109} {\bibfield  {journal}
  {\bibinfo  {journal} {\apj}\ }\textbf {\bibinfo {volume} {762}},\ \bibinfo
  {eid} {109} (\bibinfo {year} {2013})},\ \Eprint
  {http://arxiv.org/abs/1110.4372} {arXiv:1110.4372 [astro-ph.CO]} \BibitemShut
  {NoStop}%
\bibitem [{\citenamefont {{Zheng}}\ \emph {et~al.}(2007)\citenamefont
  {{Zheng}}, \citenamefont {{Coil}},\ and\ \citenamefont
  {{Zehavi}}}]{ZhengZheng2007}%
  \BibitemOpen
  \bibfield  {author} {\bibinfo {author} {\bibfnamefont {Z.}~\bibnamefont
  {{Zheng}}}, \bibinfo {author} {\bibfnamefont {A.~L.}\ \bibnamefont {{Coil}}},
  \ and\ \bibinfo {author} {\bibfnamefont {I.}~\bibnamefont {{Zehavi}}},\
  }\href {\doibase 10.1086/521074} {\bibfield  {journal} {\bibinfo  {journal}
  {\apj}\ }\textbf {\bibinfo {volume} {667}},\ \bibinfo {pages} {760} (\bibinfo
  {year} {2007})},\ \Eprint {http://arxiv.org/abs/astro-ph/0703457}
  {arXiv:astro-ph/0703457 [astro-ph]} \BibitemShut {NoStop}%
\bibitem [{\citenamefont {{Hern{\'a}ndez-Aguayo}}\ \emph
  {et~al.}(2019)\citenamefont {{Hern{\'a}ndez-Aguayo}}, \citenamefont {{Hou}},
  \citenamefont {{Li}}, \citenamefont {{Baugh}},\ and\ \citenamefont
  {{S{\'a}nchez}}}]{Cesar2019}%
  \BibitemOpen
  \bibfield  {author} {\bibinfo {author} {\bibfnamefont {C.}~\bibnamefont
  {{Hern{\'a}ndez-Aguayo}}}, \bibinfo {author} {\bibfnamefont {J.}~\bibnamefont
  {{Hou}}}, \bibinfo {author} {\bibfnamefont {B.}~\bibnamefont {{Li}}},
  \bibinfo {author} {\bibfnamefont {C.~M.}\ \bibnamefont {{Baugh}}}, \ and\
  \bibinfo {author} {\bibfnamefont {A.~G.}\ \bibnamefont {{S{\'a}nchez}}},\
  }\href {\doibase 10.1093/mnras/stz516} {\bibfield  {journal} {\bibinfo
  {journal} {\mnras}\ }\textbf {\bibinfo {volume} {485}},\ \bibinfo {pages}
  {2194} (\bibinfo {year} {2019})},\ \Eprint {http://arxiv.org/abs/1811.09197}
  {arXiv:1811.09197 [astro-ph.CO]} \BibitemShut {NoStop}%
\bibitem [{\citenamefont {{Carlstrom}}\ \emph {et~al.}(2019)\citenamefont
  {{Carlstrom}}, \citenamefont {{Abazajian}}, \citenamefont {{Addison}},
  \citenamefont {{Adshead}}, \citenamefont {{Ahmed}}, \citenamefont {{Allen}},
  \citenamefont {{Alonso}}, \citenamefont {{Alvarez}}, \citenamefont
  {{Anderson}}, \citenamefont {{Arnold}}, \citenamefont {{Baccigalupi}},
  \citenamefont {{Bailey}}, \citenamefont {{Barkats}}, \citenamefont
  {{Barron}}, \citenamefont {{Barry}}, \citenamefont {{Bartlett}},
  \citenamefont {{Basu Thakur}}, \citenamefont {{Battaglia}}, \citenamefont
  {{Baxter}}, \citenamefont {{Bean}}, \citenamefont {{Bebek}}, \citenamefont
  {{Bender}}, \citenamefont {{Benson}}, \citenamefont {{Berger}}, \citenamefont
  {{Bhimani}}, \citenamefont {{Bischoff}}, \citenamefont {{Bleem}},
  \citenamefont {{Bocquet}}, \citenamefont {{Boddy}}, \citenamefont {{Bonato}},
  \citenamefont {{Bond}}, \citenamefont {{Borrill}}, \citenamefont {{Bouchet}},
  \citenamefont {{Brown}}, \citenamefont {{Bryan}}, \citenamefont {{Burkhart}},
  \citenamefont {{Buza}}, \citenamefont {{Byrum}}, \citenamefont {{Calabrese}},
  \citenamefont {{Calafut}}, \citenamefont {{Caldwell}}, \citenamefont
  {{Carlstrom}}, \citenamefont {{Carron}}, \citenamefont {{Cecil}},
  \citenamefont {{Challinor}}, \citenamefont {{Chang}}, \citenamefont
  {{Chinone}}, \citenamefont {{Cho}}, \citenamefont {{Cooray}}, \citenamefont
  {{Crawford}}, \citenamefont {{Crites}}, \citenamefont {{Cukierman}},
  \citenamefont {{Cyr-Racine}}, \citenamefont {{de Haan}}, \citenamefont {{de
  Zotti}}, \citenamefont {{Delabrouille}}, \citenamefont {{Demarteau}},
  \citenamefont {{Devlin}}, \citenamefont {{Di Valentino}}, \citenamefont
  {{Dobbs}}, \citenamefont {{Duff}}, \citenamefont {{Duivenvoorden}},
  \citenamefont {{Dvorkin}}, \citenamefont {{Edwards}}, \citenamefont
  {{Eimer}}, \citenamefont {{Errard}}, \citenamefont {{Essinger-Hileman}},
  \citenamefont {{Fabbian}}, \citenamefont {{Feng}}, \citenamefont {{Ferraro}},
  \citenamefont {{Filippini}}, \citenamefont {{Flauger}}, \citenamefont
  {{Flaugher}}, \citenamefont {{Fraisse}}, \citenamefont {{Frolov}},
  \citenamefont {{Galitzki}}, \citenamefont {{Galli}}, \citenamefont {{Ganga}},
  \citenamefont {{Gerbino}}, \citenamefont {{Gilchriese}}, \citenamefont
  {{Gluscevic}}, \citenamefont {{Green}}, \citenamefont {{Grin}}, \citenamefont
  {{Grohs}}, \citenamefont {{Gualtieri}}, \citenamefont {{Guarino}},
  \citenamefont {{Gudmundsson}}, \citenamefont {{Habib}}, \citenamefont
  {{Haller}}, \citenamefont {{Halpern}}, \citenamefont {{Halverson}},
  \citenamefont {{Hanany}}, \citenamefont {{Harrington}}, \citenamefont
  {{Hasegawa}}, \citenamefont {{Hasselfield}}, \citenamefont {{Hazumi}},
  \citenamefont {{Heitmann}}, \citenamefont {{Henderson}}, \citenamefont
  {{Henning}}, \citenamefont {{Hill}}, \citenamefont {{Hlo{\v{z}}ek}},
  \citenamefont {{Holder}}, \citenamefont {{Holzapfel}}, \citenamefont
  {{Hubmayr}}, \citenamefont {{Huffenberger}}, \citenamefont {{Huffer}},
  \citenamefont {{Hui}}, \citenamefont {{Irwin}}, \citenamefont {{Johnson}},
  \citenamefont {{Johnstone}}, \citenamefont {{Jones}}, \citenamefont
  {{Karkare}}, \citenamefont {{Katayama}}, \citenamefont {{Kerby}},
  \citenamefont {{Kernovsky}}, \citenamefont {{Keskitalo}}, \citenamefont
  {{Kisner}}, \citenamefont {{Knox}}, \citenamefont {{Kosowsky}}, \citenamefont
  {{Kovac}}, \citenamefont {{Kovetz}}, \citenamefont {{Kuhlmann}},
  \citenamefont {{Kuo}}, \citenamefont {{Kurita}}, \citenamefont {{Kusaka}},
  \citenamefont {{Lahteenmaki}}, \citenamefont {{Lawrence}}, \citenamefont
  {{Lee}}, \citenamefont {{Lewis}}, \citenamefont {{Li}}, \citenamefont
  {{Linder}}, \citenamefont {{Loverde}}, \citenamefont {{Lowitz}},
  \citenamefont {{Madhavacheril}}, \citenamefont {{Mantz}}, \citenamefont
  {{Matsuda}}, \citenamefont {{Mauskopf}}, \citenamefont {{McMahon}},
  \citenamefont {{Meerburg}}, \citenamefont {{Melin}}, \citenamefont
  {{Meyers}}, \citenamefont {{Millea}}, \citenamefont {{Mohr}}, \citenamefont
  {{Moncelsi}}, \citenamefont {{Mroczkowski}}, \citenamefont {{Mukherjee}},
  \citenamefont {{Munchmeyer}}, \citenamefont {{Nagai}}, \citenamefont
  {{Nagy}}, \citenamefont {{Namikawa}}, \citenamefont {{Nati}}, \citenamefont
  {{Natoli}}, \citenamefont {{Negrello}}, \citenamefont {{Newburgh}},
  \citenamefont {{Niemack}}, \citenamefont {{Nishino}}, \citenamefont
  {{Nordby}}, \citenamefont {{Novosad}}, \citenamefont {{O'Connor}},
  \citenamefont {{Obied}}, \citenamefont {{Padin}}, \citenamefont {{Pand ey}},
  \citenamefont {{Partridge}}, \citenamefont {{Pierpaoli}}, \citenamefont
  {{Pogosian}}, \citenamefont {{Pryke}}, \citenamefont {{Puglisi}},
  \citenamefont {{Racine}}, \citenamefont {{Raghunathan}}, \citenamefont
  {{Rahlin}}, \citenamefont {{Rajagopalan}}, \citenamefont {{Raveri}},
  \citenamefont {{Reichanadter}}, \citenamefont {{Reichardt}}, \citenamefont
  {{Remazeilles}}, \citenamefont {{Rocha}}, \citenamefont {{Roe}},
  \citenamefont {{Roy}}, \citenamefont {{Ruhl}}, \citenamefont {{Salatino}},
  \citenamefont {{Saliwanchik}}, \citenamefont {{Schaan}}, \citenamefont
  {{Schillaci}}, \citenamefont {{Schmittfull}}, \citenamefont {{Scott}},
  \citenamefont {{Sehgal}}, \citenamefont {{Shandera}}, \citenamefont
  {{Sheehy}}, \citenamefont {{Sherwin}}, \citenamefont {{Shirokoff}},
  \citenamefont {{Simon}}, \citenamefont {{Slosar}}, \citenamefont
  {{Somerville}}, \citenamefont {{Staggs}}, \citenamefont {{Stark}},
  \citenamefont {{Stompor}}, \citenamefont {{Story}}, \citenamefont
  {{Stoughton}}, \citenamefont {{Suzuki}}, \citenamefont {{Tajima}},
  \citenamefont {{Teply}}, \citenamefont {{Thompson}}, \citenamefont
  {{Timbie}}, \citenamefont {{Tomasi}}, \citenamefont {{Treu}}, \citenamefont
  {{Tristram}}, \citenamefont {{Tucker}}, \citenamefont {{Umilta}},
  \citenamefont {{van Engelen}}, \citenamefont {{Vieira}}, \citenamefont
  {{Vieregg}}, \citenamefont {{Vogelsberger}}, \citenamefont {{Wang}},
  \citenamefont {{Watson}}, \citenamefont {{White}}, \citenamefont
  {{Whitehorn}}, \citenamefont {{Wollack}}, \citenamefont {{Wu}}, \citenamefont
  {{Xu}}, \citenamefont {{Yasini}}, \citenamefont {{Yeck}}, \citenamefont
  {{Yoon}}, \citenamefont {{Young}},\ and\ \citenamefont
  {{Zonca}}}]{CMBS42019}%
  \BibitemOpen
  \bibfield  {author} {\bibinfo {author} {\bibfnamefont {J.}~\bibnamefont
  {{Carlstrom}}}, \bibinfo {author} {\bibfnamefont {K.}~\bibnamefont
  {{Abazajian}}}, \bibinfo {author} {\bibfnamefont {G.}~\bibnamefont
  {{Addison}}}, \bibinfo {author} {\bibfnamefont {P.}~\bibnamefont
  {{Adshead}}}, \bibinfo {author} {\bibfnamefont {Z.}~\bibnamefont {{Ahmed}}},
  \bibinfo {author} {\bibfnamefont {S.~W.}\ \bibnamefont {{Allen}}}, \bibinfo
  {author} {\bibfnamefont {D.}~\bibnamefont {{Alonso}}}, \bibinfo {author}
  {\bibfnamefont {M.}~\bibnamefont {{Alvarez}}}, \bibinfo {author}
  {\bibfnamefont {A.}~\bibnamefont {{Anderson}}}, \bibinfo {author}
  {\bibfnamefont {K.~S.}\ \bibnamefont {{Arnold}}}, \bibinfo {author}
  {\bibfnamefont {C.}~\bibnamefont {{Baccigalupi}}}, \bibinfo {author}
  {\bibfnamefont {K.}~\bibnamefont {{Bailey}}}, \bibinfo {author}
  {\bibfnamefont {D.}~\bibnamefont {{Barkats}}}, \bibinfo {author}
  {\bibfnamefont {D.}~\bibnamefont {{Barron}}}, \bibinfo {author}
  {\bibfnamefont {P.~S.}\ \bibnamefont {{Barry}}}, \bibinfo {author}
  {\bibfnamefont {J.~G.}\ \bibnamefont {{Bartlett}}}, \bibinfo {author}
  {\bibfnamefont {R.}~\bibnamefont {{Basu Thakur}}}, \bibinfo {author}
  {\bibfnamefont {N.}~\bibnamefont {{Battaglia}}}, \bibinfo {author}
  {\bibfnamefont {E.}~\bibnamefont {{Baxter}}}, \bibinfo {author}
  {\bibfnamefont {R.}~\bibnamefont {{Bean}}}, \bibinfo {author} {\bibfnamefont
  {C.}~\bibnamefont {{Bebek}}}, \bibinfo {author} {\bibfnamefont {A.~N.}\
  \bibnamefont {{Bender}}}, \bibinfo {author} {\bibfnamefont {B.~A.}\
  \bibnamefont {{Benson}}}, \bibinfo {author} {\bibfnamefont {E.}~\bibnamefont
  {{Berger}}}, \bibinfo {author} {\bibfnamefont {S.}~\bibnamefont {{Bhimani}}},
  \bibinfo {author} {\bibfnamefont {C.~A.}\ \bibnamefont {{Bischoff}}},
  \bibinfo {author} {\bibfnamefont {L.}~\bibnamefont {{Bleem}}}, \bibinfo
  {author} {\bibfnamefont {S.}~\bibnamefont {{Bocquet}}}, \bibinfo {author}
  {\bibfnamefont {K.}~\bibnamefont {{Boddy}}}, \bibinfo {author} {\bibfnamefont
  {M.}~\bibnamefont {{Bonato}}}, \bibinfo {author} {\bibfnamefont {J.~R.}\
  \bibnamefont {{Bond}}}, \bibinfo {author} {\bibfnamefont {J.}~\bibnamefont
  {{Borrill}}}, \bibinfo {author} {\bibfnamefont {F.~R.}\ \bibnamefont
  {{Bouchet}}}, \bibinfo {author} {\bibfnamefont {M.~L.}\ \bibnamefont
  {{Brown}}}, \bibinfo {author} {\bibfnamefont {S.}~\bibnamefont {{Bryan}}},
  \bibinfo {author} {\bibfnamefont {B.}~\bibnamefont {{Burkhart}}}, \bibinfo
  {author} {\bibfnamefont {V.}~\bibnamefont {{Buza}}}, \bibinfo {author}
  {\bibfnamefont {K.}~\bibnamefont {{Byrum}}}, \bibinfo {author} {\bibfnamefont
  {E.}~\bibnamefont {{Calabrese}}}, \bibinfo {author} {\bibfnamefont
  {V.}~\bibnamefont {{Calafut}}}, \bibinfo {author} {\bibfnamefont
  {R.}~\bibnamefont {{Caldwell}}}, \bibinfo {author} {\bibfnamefont {J.~E.}\
  \bibnamefont {{Carlstrom}}}, \bibinfo {author} {\bibfnamefont
  {J.}~\bibnamefont {{Carron}}}, \bibinfo {author} {\bibfnamefont
  {T.}~\bibnamefont {{Cecil}}}, \bibinfo {author} {\bibfnamefont
  {A.}~\bibnamefont {{Challinor}}}, \bibinfo {author} {\bibfnamefont {C.~L.}\
  \bibnamefont {{Chang}}}, \bibinfo {author} {\bibfnamefont {Y.}~\bibnamefont
  {{Chinone}}}, \bibinfo {author} {\bibfnamefont {H.-M.~S.}\ \bibnamefont
  {{Cho}}}, \bibinfo {author} {\bibfnamefont {A.}~\bibnamefont {{Cooray}}},
  \bibinfo {author} {\bibfnamefont {T.~M.}\ \bibnamefont {{Crawford}}},
  \bibinfo {author} {\bibfnamefont {A.}~\bibnamefont {{Crites}}}, \bibinfo
  {author} {\bibfnamefont {A.}~\bibnamefont {{Cukierman}}}, \bibinfo {author}
  {\bibfnamefont {F.-Y.}\ \bibnamefont {{Cyr-Racine}}}, \bibinfo {author}
  {\bibfnamefont {T.}~\bibnamefont {{de Haan}}}, \bibinfo {author}
  {\bibfnamefont {G.}~\bibnamefont {{de Zotti}}}, \bibinfo {author}
  {\bibfnamefont {J.}~\bibnamefont {{Delabrouille}}}, \bibinfo {author}
  {\bibfnamefont {M.}~\bibnamefont {{Demarteau}}}, \bibinfo {author}
  {\bibfnamefont {M.}~\bibnamefont {{Devlin}}}, \bibinfo {author}
  {\bibfnamefont {E.}~\bibnamefont {{Di Valentino}}}, \bibinfo {author}
  {\bibfnamefont {M.}~\bibnamefont {{Dobbs}}}, \bibinfo {author} {\bibfnamefont
  {S.}~\bibnamefont {{Duff}}}, \bibinfo {author} {\bibfnamefont
  {A.}~\bibnamefont {{Duivenvoorden}}}, \bibinfo {author} {\bibfnamefont
  {C.}~\bibnamefont {{Dvorkin}}}, \bibinfo {author} {\bibfnamefont
  {W.}~\bibnamefont {{Edwards}}}, \bibinfo {author} {\bibfnamefont
  {J.}~\bibnamefont {{Eimer}}}, \bibinfo {author} {\bibfnamefont
  {J.}~\bibnamefont {{Errard}}}, \bibinfo {author} {\bibfnamefont
  {T.}~\bibnamefont {{Essinger-Hileman}}}, \bibinfo {author} {\bibfnamefont
  {G.}~\bibnamefont {{Fabbian}}}, \bibinfo {author} {\bibfnamefont
  {C.}~\bibnamefont {{Feng}}}, \bibinfo {author} {\bibfnamefont
  {S.}~\bibnamefont {{Ferraro}}}, \bibinfo {author} {\bibfnamefont {J.~P.}\
  \bibnamefont {{Filippini}}}, \bibinfo {author} {\bibfnamefont
  {R.}~\bibnamefont {{Flauger}}}, \bibinfo {author} {\bibfnamefont
  {B.}~\bibnamefont {{Flaugher}}}, \bibinfo {author} {\bibfnamefont {A.~A.}\
  \bibnamefont {{Fraisse}}}, \bibinfo {author} {\bibfnamefont {A.}~\bibnamefont
  {{Frolov}}}, \bibinfo {author} {\bibfnamefont {N.}~\bibnamefont
  {{Galitzki}}}, \bibinfo {author} {\bibfnamefont {S.}~\bibnamefont {{Galli}}},
  \bibinfo {author} {\bibfnamefont {K.}~\bibnamefont {{Ganga}}}, \bibinfo
  {author} {\bibfnamefont {M.}~\bibnamefont {{Gerbino}}}, \bibinfo {author}
  {\bibfnamefont {M.}~\bibnamefont {{Gilchriese}}}, \bibinfo {author}
  {\bibfnamefont {V.}~\bibnamefont {{Gluscevic}}}, \bibinfo {author}
  {\bibfnamefont {D.}~\bibnamefont {{Green}}}, \bibinfo {author} {\bibfnamefont
  {D.}~\bibnamefont {{Grin}}}, \bibinfo {author} {\bibfnamefont
  {E.}~\bibnamefont {{Grohs}}}, \bibinfo {author} {\bibfnamefont
  {R.}~\bibnamefont {{Gualtieri}}}, \bibinfo {author} {\bibfnamefont
  {V.}~\bibnamefont {{Guarino}}}, \bibinfo {author} {\bibfnamefont {J.~E.}\
  \bibnamefont {{Gudmundsson}}}, \bibinfo {author} {\bibfnamefont
  {S.}~\bibnamefont {{Habib}}}, \bibinfo {author} {\bibfnamefont
  {G.}~\bibnamefont {{Haller}}}, \bibinfo {author} {\bibfnamefont
  {M.}~\bibnamefont {{Halpern}}}, \bibinfo {author} {\bibfnamefont {N.~W.}\
  \bibnamefont {{Halverson}}}, \bibinfo {author} {\bibfnamefont
  {S.}~\bibnamefont {{Hanany}}}, \bibinfo {author} {\bibfnamefont
  {K.}~\bibnamefont {{Harrington}}}, \bibinfo {author} {\bibfnamefont
  {M.}~\bibnamefont {{Hasegawa}}}, \bibinfo {author} {\bibfnamefont
  {M.}~\bibnamefont {{Hasselfield}}}, \bibinfo {author} {\bibfnamefont
  {M.}~\bibnamefont {{Hazumi}}}, \bibinfo {author} {\bibfnamefont
  {K.}~\bibnamefont {{Heitmann}}}, \bibinfo {author} {\bibfnamefont
  {S.}~\bibnamefont {{Henderson}}}, \bibinfo {author} {\bibfnamefont {J.~W.}\
  \bibnamefont {{Henning}}}, \bibinfo {author} {\bibfnamefont {J.~C.}\
  \bibnamefont {{Hill}}}, \bibinfo {author} {\bibfnamefont {R.}~\bibnamefont
  {{Hlo{\v{z}}ek}}}, \bibinfo {author} {\bibfnamefont {G.}~\bibnamefont
  {{Holder}}}, \bibinfo {author} {\bibfnamefont {W.}~\bibnamefont
  {{Holzapfel}}}, \bibinfo {author} {\bibfnamefont {J.}~\bibnamefont
  {{Hubmayr}}}, \bibinfo {author} {\bibfnamefont {K.~M.}\ \bibnamefont
  {{Huffenberger}}}, \bibinfo {author} {\bibfnamefont {M.}~\bibnamefont
  {{Huffer}}}, \bibinfo {author} {\bibfnamefont {H.}~\bibnamefont {{Hui}}},
  \bibinfo {author} {\bibfnamefont {K.}~\bibnamefont {{Irwin}}}, \bibinfo
  {author} {\bibfnamefont {B.~R.}\ \bibnamefont {{Johnson}}}, \bibinfo {author}
  {\bibfnamefont {D.}~\bibnamefont {{Johnstone}}}, \bibinfo {author}
  {\bibfnamefont {W.~C.}\ \bibnamefont {{Jones}}}, \bibinfo {author}
  {\bibfnamefont {K.}~\bibnamefont {{Karkare}}}, \bibinfo {author}
  {\bibfnamefont {N.}~\bibnamefont {{Katayama}}}, \bibinfo {author}
  {\bibfnamefont {J.}~\bibnamefont {{Kerby}}}, \bibinfo {author} {\bibfnamefont
  {S.}~\bibnamefont {{Kernovsky}}}, \bibinfo {author} {\bibfnamefont
  {R.}~\bibnamefont {{Keskitalo}}}, \bibinfo {author} {\bibfnamefont
  {T.}~\bibnamefont {{Kisner}}}, \bibinfo {author} {\bibfnamefont
  {L.}~\bibnamefont {{Knox}}}, \bibinfo {author} {\bibfnamefont
  {A.}~\bibnamefont {{Kosowsky}}}, \bibinfo {author} {\bibfnamefont
  {J.}~\bibnamefont {{Kovac}}}, \bibinfo {author} {\bibfnamefont {E.~D.}\
  \bibnamefont {{Kovetz}}}, \bibinfo {author} {\bibfnamefont {S.}~\bibnamefont
  {{Kuhlmann}}}, \bibinfo {author} {\bibfnamefont {C.-l.}\ \bibnamefont
  {{Kuo}}}, \bibinfo {author} {\bibfnamefont {N.}~\bibnamefont {{Kurita}}},
  \bibinfo {author} {\bibfnamefont {A.}~\bibnamefont {{Kusaka}}}, \bibinfo
  {author} {\bibfnamefont {A.}~\bibnamefont {{Lahteenmaki}}}, \bibinfo {author}
  {\bibfnamefont {C.~R.}\ \bibnamefont {{Lawrence}}}, \bibinfo {author}
  {\bibfnamefont {A.~T.}\ \bibnamefont {{Lee}}}, \bibinfo {author}
  {\bibfnamefont {A.}~\bibnamefont {{Lewis}}}, \bibinfo {author} {\bibfnamefont
  {D.}~\bibnamefont {{Li}}}, \bibinfo {author} {\bibfnamefont {E.}~\bibnamefont
  {{Linder}}}, \bibinfo {author} {\bibfnamefont {M.}~\bibnamefont {{Loverde}}},
  \bibinfo {author} {\bibfnamefont {A.}~\bibnamefont {{Lowitz}}}, \bibinfo
  {author} {\bibfnamefont {M.~S.}\ \bibnamefont {{Madhavacheril}}}, \bibinfo
  {author} {\bibfnamefont {A.}~\bibnamefont {{Mantz}}}, \bibinfo {author}
  {\bibfnamefont {F.}~\bibnamefont {{Matsuda}}}, \bibinfo {author}
  {\bibfnamefont {P.}~\bibnamefont {{Mauskopf}}}, \bibinfo {author}
  {\bibfnamefont {J.}~\bibnamefont {{McMahon}}}, \bibinfo {author}
  {\bibfnamefont {P.~D.}\ \bibnamefont {{Meerburg}}}, \bibinfo {author}
  {\bibfnamefont {J.}~\bibnamefont {{Melin}}}, \bibinfo {author} {\bibfnamefont
  {J.}~\bibnamefont {{Meyers}}}, \bibinfo {author} {\bibfnamefont
  {M.}~\bibnamefont {{Millea}}}, \bibinfo {author} {\bibfnamefont
  {J.}~\bibnamefont {{Mohr}}}, \bibinfo {author} {\bibfnamefont
  {L.}~\bibnamefont {{Moncelsi}}}, \bibinfo {author} {\bibfnamefont
  {T.}~\bibnamefont {{Mroczkowski}}}, \bibinfo {author} {\bibfnamefont
  {S.}~\bibnamefont {{Mukherjee}}}, \bibinfo {author} {\bibfnamefont
  {M.}~\bibnamefont {{Munchmeyer}}}, \bibinfo {author} {\bibfnamefont
  {D.}~\bibnamefont {{Nagai}}}, \bibinfo {author} {\bibfnamefont
  {J.}~\bibnamefont {{Nagy}}}, \bibinfo {author} {\bibfnamefont
  {T.}~\bibnamefont {{Namikawa}}}, \bibinfo {author} {\bibfnamefont
  {F.}~\bibnamefont {{Nati}}}, \bibinfo {author} {\bibfnamefont
  {T.}~\bibnamefont {{Natoli}}}, \bibinfo {author} {\bibfnamefont
  {M.}~\bibnamefont {{Negrello}}}, \bibinfo {author} {\bibfnamefont
  {L.}~\bibnamefont {{Newburgh}}}, \bibinfo {author} {\bibfnamefont {M.~D.}\
  \bibnamefont {{Niemack}}}, \bibinfo {author} {\bibfnamefont {H.}~\bibnamefont
  {{Nishino}}}, \bibinfo {author} {\bibfnamefont {M.}~\bibnamefont {{Nordby}}},
  \bibinfo {author} {\bibfnamefont {V.}~\bibnamefont {{Novosad}}}, \bibinfo
  {author} {\bibfnamefont {P.}~\bibnamefont {{O'Connor}}}, \bibinfo {author}
  {\bibfnamefont {G.}~\bibnamefont {{Obied}}}, \bibinfo {author} {\bibfnamefont
  {S.}~\bibnamefont {{Padin}}}, \bibinfo {author} {\bibfnamefont
  {S.}~\bibnamefont {{Pand ey}}}, \bibinfo {author} {\bibfnamefont
  {B.}~\bibnamefont {{Partridge}}}, \bibinfo {author} {\bibfnamefont
  {E.}~\bibnamefont {{Pierpaoli}}}, \bibinfo {author} {\bibfnamefont
  {L.}~\bibnamefont {{Pogosian}}}, \bibinfo {author} {\bibfnamefont
  {C.}~\bibnamefont {{Pryke}}}, \bibinfo {author} {\bibfnamefont
  {G.}~\bibnamefont {{Puglisi}}}, \bibinfo {author} {\bibfnamefont
  {B.}~\bibnamefont {{Racine}}}, \bibinfo {author} {\bibfnamefont
  {S.}~\bibnamefont {{Raghunathan}}}, \bibinfo {author} {\bibfnamefont {A.~r.}\
  \bibnamefont {{Rahlin}}}, \bibinfo {author} {\bibfnamefont {S.}~\bibnamefont
  {{Rajagopalan}}}, \bibinfo {author} {\bibfnamefont {M.}~\bibnamefont
  {{Raveri}}}, \bibinfo {author} {\bibfnamefont {M.}~\bibnamefont
  {{Reichanadter}}}, \bibinfo {author} {\bibfnamefont {C.~L.}\ \bibnamefont
  {{Reichardt}}}, \bibinfo {author} {\bibfnamefont {M.}~\bibnamefont
  {{Remazeilles}}}, \bibinfo {author} {\bibfnamefont {G.}~\bibnamefont
  {{Rocha}}}, \bibinfo {author} {\bibfnamefont {N.~A.}\ \bibnamefont {{Roe}}},
  \bibinfo {author} {\bibfnamefont {A.}~\bibnamefont {{Roy}}}, \bibinfo
  {author} {\bibfnamefont {J.}~\bibnamefont {{Ruhl}}}, \bibinfo {author}
  {\bibfnamefont {M.}~\bibnamefont {{Salatino}}}, \bibinfo {author}
  {\bibfnamefont {B.}~\bibnamefont {{Saliwanchik}}}, \bibinfo {author}
  {\bibfnamefont {E.}~\bibnamefont {{Schaan}}}, \bibinfo {author}
  {\bibfnamefont {A.~r.}\ \bibnamefont {{Schillaci}}}, \bibinfo {author}
  {\bibfnamefont {M.~M.}\ \bibnamefont {{Schmittfull}}}, \bibinfo {author}
  {\bibfnamefont {D.}~\bibnamefont {{Scott}}}, \bibinfo {author} {\bibfnamefont
  {N.}~\bibnamefont {{Sehgal}}}, \bibinfo {author} {\bibfnamefont
  {S.}~\bibnamefont {{Shandera}}}, \bibinfo {author} {\bibfnamefont
  {C.}~\bibnamefont {{Sheehy}}}, \bibinfo {author} {\bibfnamefont {B.~D.}\
  \bibnamefont {{Sherwin}}}, \bibinfo {author} {\bibfnamefont {E.}~\bibnamefont
  {{Shirokoff}}}, \bibinfo {author} {\bibfnamefont {S.~M.}\ \bibnamefont
  {{Simon}}}, \bibinfo {author} {\bibfnamefont {A.}~\bibnamefont {{Slosar}}},
  \bibinfo {author} {\bibfnamefont {R.}~\bibnamefont {{Somerville}}}, \bibinfo
  {author} {\bibfnamefont {S.~T.}\ \bibnamefont {{Staggs}}}, \bibinfo {author}
  {\bibfnamefont {A.}~\bibnamefont {{Stark}}}, \bibinfo {author} {\bibfnamefont
  {R.}~\bibnamefont {{Stompor}}}, \bibinfo {author} {\bibfnamefont {K.~T.}\
  \bibnamefont {{Story}}}, \bibinfo {author} {\bibfnamefont {C.}~\bibnamefont
  {{Stoughton}}}, \bibinfo {author} {\bibfnamefont {A.}~\bibnamefont
  {{Suzuki}}}, \bibinfo {author} {\bibfnamefont {O.}~\bibnamefont {{Tajima}}},
  \bibinfo {author} {\bibfnamefont {G.~P.}\ \bibnamefont {{Teply}}}, \bibinfo
  {author} {\bibfnamefont {K.}~\bibnamefont {{Thompson}}}, \bibinfo {author}
  {\bibfnamefont {P.}~\bibnamefont {{Timbie}}}, \bibinfo {author}
  {\bibfnamefont {M.}~\bibnamefont {{Tomasi}}}, \bibinfo {author}
  {\bibfnamefont {J.~I.}\ \bibnamefont {{Treu}}}, \bibinfo {author}
  {\bibfnamefont {M.}~\bibnamefont {{Tristram}}}, \bibinfo {author}
  {\bibfnamefont {G.}~\bibnamefont {{Tucker}}}, \bibinfo {author}
  {\bibfnamefont {C.}~\bibnamefont {{Umilta}}}, \bibinfo {author}
  {\bibfnamefont {A.}~\bibnamefont {{van Engelen}}}, \bibinfo {author}
  {\bibfnamefont {J.~D.}\ \bibnamefont {{Vieira}}}, \bibinfo {author}
  {\bibfnamefont {A.~G.}\ \bibnamefont {{Vieregg}}}, \bibinfo {author}
  {\bibfnamefont {M.}~\bibnamefont {{Vogelsberger}}}, \bibinfo {author}
  {\bibfnamefont {G.}~\bibnamefont {{Wang}}}, \bibinfo {author} {\bibfnamefont
  {S.}~\bibnamefont {{Watson}}}, \bibinfo {author} {\bibfnamefont
  {M.}~\bibnamefont {{White}}}, \bibinfo {author} {\bibfnamefont
  {N.}~\bibnamefont {{Whitehorn}}}, \bibinfo {author} {\bibfnamefont {E.~J.}\
  \bibnamefont {{Wollack}}}, \bibinfo {author} {\bibfnamefont {W.~L.~K.}\
  \bibnamefont {{Wu}}}, \bibinfo {author} {\bibfnamefont {Z.}~\bibnamefont
  {{Xu}}}, \bibinfo {author} {\bibfnamefont {S.}~\bibnamefont {{Yasini}}},
  \bibinfo {author} {\bibfnamefont {J.}~\bibnamefont {{Yeck}}}, \bibinfo
  {author} {\bibfnamefont {K.~W.}\ \bibnamefont {{Yoon}}}, \bibinfo {author}
  {\bibfnamefont {E.}~\bibnamefont {{Young}}}, \ and\ \bibinfo {author}
  {\bibfnamefont {A.}~\bibnamefont {{Zonca}}},\ }in\ \href@noop {} {\emph
  {\bibinfo {booktitle} {\baas}}},\ Vol.~\bibinfo {volume} {51}\ (\bibinfo
  {year} {2019})\ p.\ \bibinfo {pages} {209},\ \Eprint
  {http://arxiv.org/abs/1908.01062} {arXiv:1908.01062 [astro-ph.IM]}
  \BibitemShut {NoStop}%
\bibitem [{\citenamefont {{Alonso}}\ \emph {et~al.}(2016)\citenamefont
  {{Alonso}}, \citenamefont {{Louis}}, \citenamefont {{Bull}},\ and\
  \citenamefont {{Ferreira}}}]{Alonso2016}%
  \BibitemOpen
  \bibfield  {author} {\bibinfo {author} {\bibfnamefont {D.}~\bibnamefont
  {{Alonso}}}, \bibinfo {author} {\bibfnamefont {T.}~\bibnamefont {{Louis}}},
  \bibinfo {author} {\bibfnamefont {P.}~\bibnamefont {{Bull}}}, \ and\ \bibinfo
  {author} {\bibfnamefont {P.~G.}\ \bibnamefont {{Ferreira}}},\ }\href
  {\doibase 10.1103/PhysRevD.94.043522} {\bibfield  {journal} {\bibinfo
  {journal} {\prd}\ }\textbf {\bibinfo {volume} {94}},\ \bibinfo {eid} {043522}
  (\bibinfo {year} {2016})},\ \Eprint {http://arxiv.org/abs/1604.01382}
  {arXiv:1604.01382 [astro-ph.CO]} \BibitemShut {NoStop}%
\bibitem [{\citenamefont {{Pan}}\ and\ \citenamefont
  {{Johnson}}(2019)}]{Pan2019}%
  \BibitemOpen
  \bibfield  {author} {\bibinfo {author} {\bibfnamefont {Z.}~\bibnamefont
  {{Pan}}}\ and\ \bibinfo {author} {\bibfnamefont {M.~C.}\ \bibnamefont
  {{Johnson}}},\ }\href {\doibase 10.1103/PhysRevD.100.083522} {\bibfield
  {journal} {\bibinfo  {journal} {\prd}\ }\textbf {\bibinfo {volume} {100}},\
  \bibinfo {eid} {083522} (\bibinfo {year} {2019})},\ \Eprint
  {http://arxiv.org/abs/1906.04208} {arXiv:1906.04208 [astro-ph.CO]}
  \BibitemShut {NoStop}%
\bibitem [{\citenamefont {{Smith}}\ \emph {et~al.}(2018)\citenamefont
  {{Smith}}, \citenamefont {{Madhavacheril}}, \citenamefont {{M{\"u}nchmeyer}},
  \citenamefont {{Ferraro}}, \citenamefont {{Giri}},\ and\ \citenamefont
  {{Johnson}}}]{Smith2018}%
  \BibitemOpen
  \bibfield  {author} {\bibinfo {author} {\bibfnamefont {K.~M.}\ \bibnamefont
  {{Smith}}}, \bibinfo {author} {\bibfnamefont {M.~S.}\ \bibnamefont
  {{Madhavacheril}}}, \bibinfo {author} {\bibfnamefont {M.}~\bibnamefont
  {{M{\"u}nchmeyer}}}, \bibinfo {author} {\bibfnamefont {S.}~\bibnamefont
  {{Ferraro}}}, \bibinfo {author} {\bibfnamefont {U.}~\bibnamefont {{Giri}}}, \
  and\ \bibinfo {author} {\bibfnamefont {M.~C.}\ \bibnamefont {{Johnson}}},\
  }\href@noop {} {\bibfield  {journal} {\bibinfo  {journal} {arXiv e-prints}\
  ,\ \bibinfo {eid} {arXiv:1810.13423}} (\bibinfo {year} {2018})},\ \Eprint
  {http://arxiv.org/abs/1810.13423} {arXiv:1810.13423 [astro-ph.CO]}
  \BibitemShut {NoStop}%
\bibitem [{\citenamefont {{Zheng}}\ \emph {et~al.}(2018)\citenamefont
  {{Zheng}}, \citenamefont {{Zhao}}, \citenamefont {{Li}}, \citenamefont
  {{Wang}}, \citenamefont {{Chuang}}, \citenamefont {{Kitaura}},\ and\
  \citenamefont {{Rodriguez-Torres}}}]{ZhengJ2018}%
  \BibitemOpen
  \bibfield  {author} {\bibinfo {author} {\bibfnamefont {J.}~\bibnamefont
  {{Zheng}}}, \bibinfo {author} {\bibfnamefont {G.-B.}\ \bibnamefont {{Zhao}}},
  \bibinfo {author} {\bibfnamefont {J.}~\bibnamefont {{Li}}}, \bibinfo {author}
  {\bibfnamefont {Y.}~\bibnamefont {{Wang}}}, \bibinfo {author} {\bibfnamefont
  {C.-H.}\ \bibnamefont {{Chuang}}}, \bibinfo {author} {\bibfnamefont {F.-S.}\
  \bibnamefont {{Kitaura}}}, \ and\ \bibinfo {author} {\bibfnamefont
  {S.}~\bibnamefont {{Rodriguez-Torres}}},\ }\href@noop {} {\bibfield
  {journal} {\bibinfo  {journal} {ArXiv e-prints}\ } (\bibinfo {year}
  {2018})},\ \Eprint {http://arxiv.org/abs/1806.01920} {arXiv:1806.01920}
  \BibitemShut {NoStop}%
\bibitem [{\citenamefont {{Planck Collaboration}}\ \emph
  {et~al.}(2015)\citenamefont {{Planck Collaboration}}, \citenamefont {{Ade}},
  \citenamefont {{Aghanim}}, \citenamefont {{Arnaud}}, \citenamefont
  {{Ashdown}}, \citenamefont {{Aumont}}, \citenamefont {{Baccigalupi}},
  \citenamefont {{Banday}}, \citenamefont {{Barreiro}}, \citenamefont
  {{Bartlett}},\ and\ \citenamefont {et~al.}}]{PLANK2015}%
  \BibitemOpen
  \bibfield  {author} {\bibinfo {author} {\bibnamefont {{Planck
  Collaboration}}}, \bibinfo {author} {\bibfnamefont {P.~A.~R.}\ \bibnamefont
  {{Ade}}}, \bibinfo {author} {\bibfnamefont {N.}~\bibnamefont {{Aghanim}}},
  \bibinfo {author} {\bibfnamefont {M.}~\bibnamefont {{Arnaud}}}, \bibinfo
  {author} {\bibfnamefont {M.}~\bibnamefont {{Ashdown}}}, \bibinfo {author}
  {\bibfnamefont {J.}~\bibnamefont {{Aumont}}}, \bibinfo {author}
  {\bibfnamefont {C.}~\bibnamefont {{Baccigalupi}}}, \bibinfo {author}
  {\bibfnamefont {A.~J.}\ \bibnamefont {{Banday}}}, \bibinfo {author}
  {\bibfnamefont {R.~B.}\ \bibnamefont {{Barreiro}}}, \bibinfo {author}
  {\bibfnamefont {J.~G.}\ \bibnamefont {{Bartlett}}}, \ and\ \bibinfo {author}
  {\bibnamefont {et~al.}},\ }\href@noop {} {\bibfield  {journal} {\bibinfo
  {journal} {ArXiv e-prints}\ } (\bibinfo {year} {2015})},\ \Eprint
  {http://arxiv.org/abs/1502.01589} {arXiv:1502.01589} \BibitemShut {NoStop}%
\end{thebibliography}%
\end{document}